# The Existence Regions for Transcomplex Recombination of Heavy Ions. Two-Dimensional Recombination Regions and Surroundings Indices

**V. M. Akimov[a], V. M. Azriel'[a], E. V. Ermolova[a], D. B. Kabanov[a], L. I. Kolesnikova[a], L. Yu. Rusin[a,]*, and M. B. Sevryuk[a]**

[a] *Semënov Federal Research Center for Chemical Physics, Russian Academy of Sciences, Moscow, 119334 Russia*

**e-mail: rusin@chph.ras.ru*

**Abstract**—In this paper, we examine transcomplex recombination $RCs^+ + Br^- \rightarrow CsBr + R$ (R = Kr, Xe, Hg) for collision energies from 0.1 to 2.5 eV. Within the framework of the quasiclassical trajectory method on semiempirical diabatic potential energy surfaces, we have considered the existence regions for recombination in the space of kinematic parameters. Two-dimensional "slices" of the existence regions for recombination (on the sphere where the coordinates are the orientation angles of the initial direction of the axis of the ionic complex $RCs^+$) sometimes exhibit a chaotic structure. We have proposed a quantitative measure of chaoticity of two-dimensional existence regions for recombination (on the basis of the combinatorics of matrices of zeros and ones) and have presented examples of chaotic regions. We have introduced the surroundings index of recombinative trajectories and have explored its connection with the integration time of the trajectory and with the total internal energy of the product molecule CsBr.

# 1. INTRODUCTION

Reactions of collision induced dissociation (CID) of neutral molecules and reverse reactions of recombination of atoms, ions, and radicals are a key component of very many processes occurring in various gaseous and plasma media, whether they are plasma of the interstellar space, the solar corona, the upper layers of the Earth atmosphere, or plasmas of artificial origin. For instance, of the 196 reactions identified in the work [1] as important for combustion chemistry, about half are CID or recombination reactions [2, 3]. The concentration of ions in plasmas is determined by the competition of reactions of CID to ionic products, on the one hand, and reactions of ion-ion recombination and mutual neutralization, on the other hand. Ion-ion recombination is widely used in various vacuum-plasma technologies, and its kinetics is the subject of numerous studies [4–12]. As a typical example of practical application of ion-ion recombination, one can mention the population of light-emitting states of rare gas monohalide excimer lasers as a result of recombination of rare gas cations and halide anions [13–15] (see also Section XI.3.2.4 of Volume IV of the encyclopedia [8]).

For already more than two decades, our group has been studying the dynamics of recombination of cesium cations and halide (fluoride, bromide, or iodide) anions by the quasiclassical trajectory method on semiempirical diabatic potential energy surfaces (PESs) [16–37] as well as within the framework of the hard sphere model [38–40].

As is well known, there exist three basic mechanisms of recombination of atomic particles (atoms or monoatomic ions) A and B [2, 3, 8, 10–12, 16–18, 20, 25, 41–43]. The first mechanism is direct three-body recombination A + B + R $\rightarrow$ AB + R, where R is a neutral atom which takes away excess energy of the recombining pair (A, B) and is called the third body. The second mechanism is the so-called bound complex mechanism also known as the radical-molecule complex mechanism, the exchange mechanism, or the chaperone mechanism. This indirect mechanism which includes the formation of the excited complex

RA* or RB* consists of three successive reactions A + R → RA*, RA* + R → RA + R, RA + B → AB + R or B + R → RB*, RB* + R → RB + R, RB + A → AB + R. We will call the final stages of the bound complex mechanism, i.e., the reactions RA + B → AB + R and RB + A → AB + R *transcomplex* recombination (that is, recombination occurring via a complex) of the atomic particles A and B. The third mechanism, namely, the Lindemann (or Lindemann–Hinshelwood) energy transfer mechanism, is also indirect and has the form A + B → AB*, AB* + R → AB + R with the excited aggregate AB* as an intermediate product.

In the works [16–40], we explored direct three-body recombination reactions [16–33, 35, 37–39]

$$Cs^+ + X^- + R \rightarrow CsX + R \quad (1)$$

and transcomplex recombination reactions [31, 34, 36, 40]

$$RCs^+ + X^- \rightarrow CsX + R, \quad (2)$$

where $X^-$ is the halide anion $F^-$, $Br^-$, $I^-$ and R is the rare gas atom Ar, Kr, Xe or the mercury atom. Transcomplex recombination (2) is the final stage of ion-ion recombination via the bound complex mechanism [32, 34, 36, 37, 40]: $Cs^+ + R \rightarrow (RCs^+)^*$, $(RCs^+)^* + R \rightarrow RCs^+ + R$, $RCs^+ + X^- \rightarrow CsX + R$. The dynamics of the reactions (1) and (2) is very different [31].

Obtaining experimental dynamical information (in crossed molecular beams) on the processes (1), and also to some extent on the processes (2), is associated with significant difficulties [16–18, 26, 30, 32, 33, 35–37, 40]. Therefore, the only means of studying the dynamics of the reactions (1) and (2) is computer simulation. Since all the particles involved in these reactions are sufficiently heavy and include large numbers of electrons (even the system $Cs^+$–$F^-$–Ar already contains 82 electrons), the quasiclassical trajectory method is the only possible and at the same time adequate option for simulating the processes (1) and (2). On the other hand, the dynamics of the reverse CID reactions

$$CsX + R \rightarrow Cs^+ + X^- + R, \;\; RCs^+ + X^- \quad (3)$$

(with the same $X^-$ and R) has been investigated in molecular beam experiments in sufficient detail, and within those studies, semiempirical diabatic PESs have been constructed that ensure quantitative reproduction of the experimental data in quasiclassical trajectory simulation (see the surveys [30, 44–47], the works [16–18, 26, 32, 48–50], and the annotated bibliography [51]). Being based on the microscopic reversibility principle, we have employed these PESs for trajectory simulation of the reactions (1) and (2) [16–37]. Note that the crucial factor that allows one to describe the processes (1)–(3) as a motion of the three particles $Cs^+$, $X^-$, R on a single diabatic PES is the fact that all the cesium halides dissociate completely to ions rather than to neutral atoms [52].

The third a priori possible CID channel $CsX + R \rightarrow RX^- + Cs^+$ was observed in experiments with crossed molecular beams for the system CsI + Xe only [48, 49, 51]. Therefore, we have not considered transcomplex recombination reactions $RX^- + Cs^+ \rightarrow CsX + R$. One may also interpret these hypothetical reactions as the final stage of ion-ion recombination via the bound complex mechanism: $X^- + R \rightarrow (RX^-)^*$, $(RX^-)^* + R \rightarrow RX^- + R$, $RX^- + Cs^+ \rightarrow CsX + R$.

In our studies of the recombination reactions (1) and (2), we computed the “standard” fundamental statistical dynamical characteristics of the process such as the excitation functions, the opacity functions, the distributions of the vibrational and rotational energies of the salt molecules CsX, or the distributions of the relative kinetic energy of the products CsX and R. Besides that, we also explored *detailed dynamics* of the process (i.e., we carried out step-by-step analysis of individual trajectories [16, 18, 19, 21–23, 25–27, 33, 36]), we determined the so-called *effectivity functions* of the third body R as of an acceptor of excess energy of the recombining ionic pair in direct three-body recombination (1) [20, 21, 24, 26–28, 30], and we considered the *existence regions* (or, to be more precise, the occurrence regions) for direct three-body recombination (1) in the space of kinematic parameters, that is, the quantities that constitute the initial conditions of trajectories [20, 26, 28, 37]. Note that we only determined the existence regions for direct three-body recombination (1) in the case where the encounter of the ions is central (head-on) and the so-called delay parameter vanishes. The subject of research in the present paper is the existence regions for transcomplex recombination (2) in the space of kinematic parameters. Just as in our previous works where transcomplex recombination (2) was examined by the quasiclassical trajectory method [31, 34, 36], we confine ourselves with the case where the ionic complex $RCs^+$ at the beginning of a trajectory is in the ground state ($v = j = 0$).

As in the works [20, 26, 28, 37] where direct three-body recombination (1) is dealt with, we consider the existence regions for recombination (2) for $X^-$ = $Br^-$ and R = Kr, Xe, Hg only. As in the paper [37], for brevity, we will just speak of the “recombination regions” in place of “the existence regions for recombination”.

The paper is organized as follows. After this introduction, we describe in Section 2 the kinematic parameters we employ for transcomplex recombination (2), and we outline the calculation procedure in Section 3. The calculation results are presented in Sections 4–6. In Section 4, we consider two-dimensional “slices” of recombination regions, we construct the characteristic matrices of these two-dimensional regions, and we define the complexity of such matrices. This complexity is a quantitative measure of chaoticity of two-dimensional recombination regions. In Section 5, we define the surroundings index of a recombinative trajectory and discuss the connection of this index with the integration time of the trajectory and with the stabilization depth of the recombination product CsBr. Section 6 is devoted to the distributions of the orientation angles of the initial direction of the axis of the $RCs^+$ complex. The remarks of Section 7 conclude the article.

## 2. INITIAL CONDITIONS OF THE TRAJECTORIES

In quasiclassical trajectory simulation of direct three-body recombination (1) in the case of central encounters of the ions with zero delay parameter, the initial conditions of each trajectory are specified by five kinematic parameters: the ion encounter energy $E_{\mathrm{i}}$, the third body energy $E_{\mathrm{R}}$, the impact parameter $b_{\mathrm{R}}$ of the third body R, as well as the polar angle (Θ) and the azimuthal angle (Φ) of the vector of the initial relative velocity of the ions ($0 \le \Theta \le 180°$, $0 \le \Phi < 360°$) [16–20, 22–24, 27, 28, 30, 31, 37]. Analogously, in quasiclassical trajectory simulation of transcomplex recombination (2) in the case of the ground state of the ionic complex $RCs^{+}$ ($v = j = 0$) at $t = 0$, the initial conditions of each trajectory are also specified by five kinematic parameters. As those parameters, one can use the collision energy $E_{\mathrm{col}}$, the impact parameter $b$, the quantity $\zeta = \kappa r_0$, as well as the polar angle (θ) and the azimuthal angle (φ) of the initial direction of the axis of the $RCs^{+}$ complex ($0 \le \theta \le 180°$, $0 \le \varphi < 360°$) [31, 34, 36]. Here $r_0$ is the initial internuclear distance between the particles R and $Cs^{+}$ while $\kappa = \pm 1$, to be more precise, $\zeta = r_0$ if the particles R and $Cs^{+}$ move away from each other at $t = 0$, and $\zeta = -r_0$ if the particles R and $Cs^{+}$ come closer together at $t = 0$.

We will define the precise meaning of the angles θ and φ according to the following convention [36]. The components of any vector $\mathbf{a}$ will be denoted by $a_1, a_2, a_3$. Let $\mathbf{q}(t)$ be the vector drawn from R to $Cs^{+}$, let $\mathbf{Q}(t)$ be the vector drawn from $X^{-}$ to the center of mass of the pair $Cs^{+}$–R, and let $\mathbf{p}(t)$ and $\mathbf{P}(t)$ be the momenta conjugate to the vectors $\mathbf{q}(t)$ and $\mathbf{Q}(t)$. Choose the coordinate frame $xyz$ in which at the beginning of a trajectory,

$$Q_1(0) = (d_{\mathrm{i}}^2 - b^2)^{1/2}, \quad Q_2(0) = b, \quad Q_3(0) = 0,$$

$$P_1(0) = -(2\mu E_{\mathrm{col}})^{1/2}, \quad P_2(0) = P_3(0) = 0, \tag{4}$$

where μ is the reduced mass of the $X^-$ ion and the $RCs^+$ complex while $d_i$ is the initial distance between the $X^-$ ion and the center of mass of the complex. Then in this coordinate frame,

$$q_1(0) = r_0 \cos\theta, \quad q_2(0) = r_0 \sin\theta\cos\varphi, \quad q_3(0) = r_0 \sin\theta\sin\varphi.$$

We always set the initial distance $d_i$ between the $X^-$ ion and the center of mass of the $RCs^+$ complex to be equal to 250 a.u., as in the papers [31, 34, 36].

One sees from the formulas (4) that by $E_{col}$ and $b$, we mean the energy of the relative motion of the $X^-$ ion and the $RCs^+$ complex and the corresponding impact parameter at the initial time instant $t = 0$, when the distance between the centers of mass of the reactants is equal to $d_i = 250$ a.u. rather than to infinity. An analogous convention was also adopted in our preceding papers [31, 34, 36] where we carried out quasiclassical trajectory simulation of transcomplex recombination (2). The initial internuclear distance $d_{st}$ between the particles $X^-$ and $Cs^+$ is close to $d_i$, so that the initial Coulomb potential $\eta_{st} = -V_{st} = -1/d_{st}$ of the interaction of the reactants $X^-$ and $RCs^+$ (in atomic units) is close to $\eta_i = -V_i = -1/d_i$ ($V_i = 0.004\,E_h = 0.108846$ eV). The subscript "st" here refers to the word "start". Consequently, the "genuine" collision energy $E_{-\infty}$ in our calculations, i.e., the energy of the relative motion of the reactants at infinity, is smaller than $E_{col} > V_{st}$ and equal to $E_{col} - V_{st}$, whereas the "genuine" impact parameter $b_{-\infty}$, on the contrary, is larger than $b < d_i$ and equal to $(E_{col}/E_{-\infty})^{1/2} b$. In the case where $E_{col} \le V_{st}$, the "genuine" collision energy $E_{-\infty}$ is not defined at all.

Note that for $\theta = 0$ and $\theta = 180°$ (i.e., for the minimal and maximal possible values of the polar angle θ), the initial conditions of a trajectory are independent of the azimuthal angle φ. Besides that, for $b = 0$, trajectories that differ only in the value of the φ angle are taken to each other by rotations about the $x$ axis and therefore possess identical dynamical characteristics. Finally, trajectories that differ only in the values $\varphi = \varphi^0$ and $\varphi = 360° - \varphi^0$ are symmetric to each other with respect to the $xy$ plane and possess identical characteristics as well. The case where $b = 0$ and $\theta = 0$ corresponds to a collinear collision configuration with an impact of the $X^-$ ion on the R atom. The case where $b = 0$ and $\theta = 180°$ corresponds to a collinear collision configuration with an impact of the $X^-$ ion on the $Cs^+$ ion.

## 3. CALCULATION PROCEDURE

We used the same semiempirical diabatic PESs as in all our previous works [16–31, 37] where we studied direct three-body recombination $Cs^+ + Br^- + R$ and transcomplex recombination $RCs^+ + Br^-$ with R = Kr, Xe, Hg by the quasiclassical trajectory method. Analytic expressions for these PESs together with the values of all the parameters are given in the works [17, 18, 21, 25–27, 29] (see also a detailed discussion of the choice of these parameters in the works [50, 53]).

To each pair of quantum numbers $(v, j)$ of the $RCs^+$ complex, there corresponds a certain interval of admissible values of the $r_0$ distance, and to each value of $r_0$ in this interval, a certain initial momentum $p_0 = |\mathbf{p}(0)|$ is associated. We determined these characteristics within the framework of the same procedure as in all our previous papers where we considered CID of cesium halides and transcomplex recombination (2) [16, 30, 31, 34, 36, 46, 47, 50]. Let us describe this procedure in more detail in relation to the ground state $v = j = 0$ of the $RCs^+$ complex. Let $r = |\mathbf{q}|$ be the internuclear distance between the particles R and $Cs^+$, and let $U(r)$ be the interaction potential between these particles with an equilibrium distance $r_{mc}$ and potential well depth $D_c = -U(r_{mc})$; here the subscript $c$ refers to the word "complex". To determine the zero-point energy $\varepsilon_{0c}$ of the complex, we approximated the potential $U(r)$ near the potential well minimum by the harmonic oscillator potential $U''(r_{mc})(r - r_{mc})^2 / 2 - D_c$, so that $\varepsilon_{0c} = [U''(r_{mc}) / \mu_c]^{1/2} / 2$ in atomic units, where $\mu_c$ is the reduced mass of the R atom and the $Cs^+$ ion. Now the left end $r_L$ and the right end $r_R$ of the interval of admissible values of the $r_0$ distance can be computed from the equations $U(r_L) = U(r_R) = \varepsilon_{0c} - D_c$, and to each value of $r_0$ in the interval $r_L < r_0 < r_R$, there corresponds the momentum $p_0$ which can be determined by the equation $p_0^2/(2\mu_c) + U(r_0) = \varepsilon_{0c} - D_c$.

As in all our previous works [16–37, 46, 47, 50], the Hamiltonian equations of motion were integrated in the present paper by the sixth order Adams–Bashforth method, while the first five integration steps for this method after having selected the kinematic parameters were carried out by the fourth order Runge–Kutta procedure. The integration step length was always set to be equal to 50 a.u. of time which ensured good conservation of the total energy and the total angular momentum of the system over a whole trajectory.

The interaction $RCs^+ + X^-$ can a priori proceed along the five channels with the products $RCs^+ + X^-$ (non-chemical scattering), $Cs^+ + X^- + R$ (dissociation of the ionic complex $RCs^+$), CsX + R (recombination, this is the channel that interests us most), $RX^- + Cs^+$ (exchange), and CsXR (the formation of the three-particle complex). The criteria of the occurrence of the first four channels in our calculations were as follows. Let A and B be any two of the particles $Cs^+$, $Br^-$, R. Then, if we deal with a channel where these particles form the bound compound AB, then the internuclear distance $r_{\mathrm{A-B}}$ between them should be smaller than 30 a.u. whereas the total internal energy $E_{\mathrm{A-B}}$ of the A–B pair should be negative. On the other hand, if we deal with a channel where the particles A and B do not form a bound compound, then the inequalities $r_{\mathrm{A-B}} > 250$ a.u. and $E_{\mathrm{A-B}} > 0$ should be satisfied. For instance, the recombination criterion consisted in the inequalities

$$r_{\mathrm{Cs^+-Br^-}} < 30 \text{ a.u.}, \quad r_{\mathrm{Cs^+-R}} > 250 \text{ a.u.}, \quad r_{\mathrm{Br^--R}} > 250 \text{ a.u.},$$

$$E_{\mathrm{Cs^+-Br^-}} < 0, \quad E_{\mathrm{Cs^+-R}} > 0, \quad E_{\mathrm{Br^--R}} > 0.$$

Besides that, the criterion of non-chemical scattering also included the inequality $Q_1 < 0$ (this inequality prevents the criterion from being met at the very beginning of the trajectory). The criterion of the occurrence of the fifth channel (the formation of the CsBrR complex) consisted in that all the three distances $r_{\mathrm{Cs^+-Br^-}}$, $r_{\mathrm{Cs^+-R}}$, $r_{\mathrm{Br^--R}}$ are smaller than 30 a.u. and $t = 5\times10^6$ a.u. of time.

We checked whether these criteria were satisfied after each integration step. If the criterion of the occurrence of any of the five channels was met, we assigned the trajectory to that channel and stopped integrating it. On the other hand, if after $T_{\mathrm{max}} = 2.5\times10^7$ a.u. of time none of the criteria turned out to be satisfied, the integration of the trajectory was also terminated.

In the case of a recombinative trajectory (i.e., a trajectory corresponding to recombination), the total internal energy of the CsBr molecule formed was set to be equal to $E_{\mathrm{mol}} = E_{\mathrm{Cs^+-Br^-}} + D$ where $D = 4.39196$ eV is the potential well depth of the $Cs^+$–$Br^-$ interaction potential. Besides that, we separately determined the vibrational energy $E_{\mathrm{vib}}$ of the CsBr product (this energy was also measured from the minimum $\varepsilon_m = -D$ of the potential well of the $Cs^+$–$Br^-$ interaction potential) and the rotational energy $E_{\mathrm{rot}} = E_{\mathrm{mol}} - E_{\mathrm{vib}}$ as well as the energy $E_{\mathrm{fly}}$ of the flying apart of the products CsBr and R. We recorded the values of the energies $E_{\mathrm{mol}}$, $E_{\mathrm{vib}}$, $E_{\mathrm{rot}}$, and $E_{\mathrm{fly}}$ for each recombinative trajectory in a special file in units of eV with three digits after the decimal point.

For each atom R = Kr, Xe, Hg, we considered trajectories for seven different values $r_{0k}$ ($1 \le k \le 7$) of the initial internuclear distance $r_0$ between the particles R and $Cs^+$ in the $RCs^+$ complex. These values were chosen as follows. Let $\delta = 0.005$ a.u. Divide the interval $r_L + \delta \le r_0 \le r_R - \delta$ into six equal parts (recall that $r_L < r_0 < r_R$ is the interval of admissible values of the $r_0$ distance corresponding to the ground state of the complex). Then $r_{0k}$ are the ends and the division points of the interval $r_L + \delta \le r_0 \le r_R - \delta$ rounded to three digits after the decimal point in atomic units. For R = Kr, the values $r_{0k}$ are equal to 6.452, 6.522, 6.592, 6.663, 6.733, 6.803, and 6.873 a.u. For R = Xe, the values $r_{0k}$ are

7.364, 7.438, 7.513, 7.587, 7.661, 7.735, and 7.81 a.u. Finally, for R = Hg, the values $r_{0k}$ turn out to be equal to 6.692, 6.746, 6.801, 6.855, 6.909, 6.964, and 7.018 a.u. It will be suitable for us to assume that $r_{01} < r_{02} < \cdots < r_{07}$.

For each of the three atoms R = Kr, Xe, Hg, we integrated the equations of motion for nine collision energies $E_{\text{col}} = 0.1$, 0.2, 0.3, 0.5, 0.75, 1, 1.5, 2, and 2.5 eV, for the impact parameter $b$ from 0 to 125 a.u. (at $E_{\text{col}} = 0.1$ eV) or from 0 to 100 a.u. (for the larger values of $E_{\text{col}}$) with a step of 1 a.u., for 14 values $\zeta = \pm r_{0k}$ $(1 \le k \le 7)$ of the kinematic parameter $\zeta = \kappa r_0$, for the polar angle θ from 0 to $180°$ with a step of $1°$, and for the azimuthal angle φ from 0 to $180°$ with a step of $1°$ (recall that trajectories that differ only in the values $\varphi = \varphi^0$ and $\varphi = 360° - \varphi^0$ possess identical dynamical characteristics, so that it is sufficient to consider angles φ that do not exceed $180°$). Thus, for each of the atoms R, a total of $(126 + 8 \times 101) \times 14 \times 181^2 = 428382836$ trajectories were integrated. Not all these trajectories are geometrically distinct because for $\theta = 0$ and $\theta = 180°$, the initial conditions of a trajectory are independent of the φ angle. In the sequel, it will be appropriate for us to assume that the 14 values of the kinematic parameter $\zeta = \kappa r_{0k}$ we consider are arranged in the ascending order of the quantity $3k - \kappa$, i.e., in the following way:

$$r_{01},\ -r_{01},\ r_{02},\ -r_{02},\ \ldots,\ r_{07},\ -r_{07}. \tag{5}$$

It is obvious that with a high accuracy, the energy balance equation of a recombinative trajectory has the form

$$E_{\text{col}} - V_{\text{st}} - D_c + \varepsilon_{0c} = -D + E_{\text{mol}} + E_{\text{fly}}. \tag{6}$$

Recall that here $V_{\text{st}}$ is the initial Coulomb potential of the interaction of the ions $X^-$ and $Cs^+$ taken with the opposite sign, $D_c$ is the potential well depth of the interaction potential in the $RCs^+$ complex, $\varepsilon_{0c}$ is the zero-point energy of the complex, $D = 4.39196$ eV is the potential well depth of the $Cs^+$–$Br^-$ interaction potential, $E_{\text{mol}} < D$ is the total internal energy of the CsBr molecule formed, and $E_{\text{fly}}$ is the energy of the flying apart of the products. In atomic units, $V_{\rm st} = 1/d_{\rm st}$ where $d_{\rm st}$ is the initial internuclear distance between the ions $X^-$ and $Cs^+$. The a priori maximal possible value of $E_{\rm mol}$ for a given collision energy $E_{\rm col}$ is equal to the smaller of the two quantities $D$ and $E_{\rm col} + D - D_c + \varepsilon_{0c} - V_{\rm st,min}$ where $V_{\rm st,min}$ is the minimal possible value of the energy $V_{\rm st}$. Obviously, the initial distance from the center of mass of the $RCs^+$ complex to the nucleus of the $Cs^+$ ion is equal to $m_{\rm R} r_0 /(m_{\rm R} + m_{\rm Cs^+})$. Consequently, in atomic units,

$$V_{\rm st,min} = \frac{1}{d_{\rm st,max}} = \frac{1}{d_{\rm i} + m_{\rm R} r_{07} /(m_{\rm R} + m_{\rm Cs^+})},$$

where $d_{\rm st,max}$ is the maximal possible value of the distance $d_{\rm st}$ in our calculation procedure, $d_{\rm i} = 250$ a.u. is the initial distance between the centers of mass of the reactants $X^-$ and $RCs^+$, while $m_{\rm R}$ and $m_{\rm Cs^+}$ are the masses of the particles R and $Cs^+$, respectively.

In Table 1, we present the energy characteristics of the transcomplex recombination reactions

$$RCs^+ + Br^- \to CsBr + R \qquad (7)$$

(R = Kr, Xe, Hg), i.e., the values of the energies $D_c$, $\varepsilon_{0c}$, $V_{\rm st,min}$, and $D - D_c + \varepsilon_{0c} - V_{\rm st,min}$ for each atom R. In this table, we also point out the value of the energy $D - D_c + \varepsilon_{0c} - V_{\rm i}$ where $V_{\rm i} = 1/d_{\rm i} = 0.108846$ eV.

## 4. TWO-DIMENSIONAL RECOMBINATION REGIONS AND THE COMPLEXITY OF THEIR CHARACTERISTIC MATRICES

The present work is mainly devoted to the analysis of two-dimensional regions of transcomplex recombination (7) on the *orientation sphere*, i.e., on the sphere of unit radius where the coordinates are the orientation angles θ and φ of the initial direction of the axis of the $RCs^+$ complex. These two-dimensional recombination regions (two-dimensional “slices”) are obtained by fixing the values of the kinematic parameters $E_{col}$, $b$, and $\zeta = \kappa r_{0k}$. In the paper [37], for direct three-body recombination

$$Cs^+ + Br^- + R \rightarrow CsBr + R \qquad (8)$$

(R = Kr, Xe, Hg), we considered two-dimensional recombination regions on the sphere of unit radius with the coordinates $(\Theta, \Phi)$ for fixed values of the kinematic parameters $E_i$, $E_R$, and $b_R$. Testing the angles Θ and Φ was carried out with a step of $1°$, like testing the angles θ and φ in the present work. As was shown in [37], the two-dimensional regions of direct three-body recombination (8) can be of a rather complicated shape; for example, they can consist of a large number (up to 10 or more) of connected components, and these components themselves can be not simply connected, i.e., they can exhibit “holes”, and in these “holes” there can be “secondary islets” of the existence of recombination. However, “locally” the two-dimensional regions of recombination (8) are almost always structured simply: they are a collection of a certain number of *solid* “domains” bounded by deformed circles.

On the other hand, at low collision energies $E_{col}$, the two-dimensional regions of transcomplex recombination (7) can exhibit *chaotic structure* on some pieces of the orientation sphere. On such pieces, both the points $(\theta, \varphi)$ corresponding to recombination and the points $(\theta, \varphi)$ where recombination does not happen are located more or less densely. Moreover, in this case, both the recombinative and non-recombinative points $(\theta, \varphi)$ are either randomly scattered across the piece of chaos or organized into alternating curved stripes of varying thickness.

The chaoticity degree of a two-dimensional recombination region can be estimated quantitatively as follows. Consider the square matrix $A = (a_{ij})$ of order 181 ($1 \le i \le 181$, $1 \le j \le 181$) whose entry $a_{ij}$ is equal to one if the point with coordinates ($\theta = j - 1$ degrees, $\varphi = i - 1$ degrees) on the orientation sphere corresponds to recombination and is equal to zero otherwise. One can call the matrix $A$ the *characteristic matrix* of the two-dimensional recombination region (for the case where the step along both the angles $\theta$ and $\varphi$ is equal to $1°$). The matrix $A$ is *logical* (other terms being binary matrix, relation matrix, Boolean matrix, $(0,1)$-matrix, matrix over the field $\mathbf{Z}_2=\{0;1\}$ with two elements), i.e., each of its entries is equal to either zero or one [54–57]. As a measure of chaoticity of a two-dimensional recombination region, one can utilize the complexity of the characteristic logical matrix. The literature contains many different definitions of the complexity of arbitrary logical matrices $A=(a_{ij})$ (not necessarily of square matrices of order 181), or, equivalently, definitions of the complexity of sign matrices (each entry of which is equal to either 1 or $-1$) [55–57]. Some of those definitions are quite “tricky”.

In the present paper, we employ the following definition of complexity of a logical matrix $A=(a_{ij})$ of an arbitrary dimension $m\times n$ ($1\le i\le m$, $1\le j\le n$). As a rule, an entry $a_{ij}$ of a matrix $A$ possesses four *neighboring* entries $a_{i,j-1}$, $a_{i-1,j}$, $a_{i,j+1}$, and $a_{i+1,j}$. For entries $a_{ij}$ “on the border” of the matrix, the number of neighboring entries drops to three or two (or even to one or zero, if $m=1$ or $n=1$). Define the *complexity* $C(A)$ of a matrix $A$ as the number of entries $a_{ij}$ that have at least one neighboring entry not equal to $a_{ij}$ (one may call such a neighboring entry a *discordant* entry). If a matrix $A=(a_{ij})$ is the characteristic matrix of a two-dimensional recombination region, then all the entries $a_{ij}$ that have at least one discordant neighboring entry constitute, in a certain sense (close to the topological sense), the “boundary” of the recombination region in question.

We have been unable to find such a definition of complexity of logical matrices in the literature, but Professor Nathan Linial (from the Hebrew University of Jerusalem) pointed out to us that this definition is reminiscent of the notion of sensitivity of a Boolean function $f:\mathbf{Z}_2^N\to\mathbf{Z}_2$ [58–62]. Let $w=(w_1,w_2,\dots,w_N)$ be an arbitrary vector in $\mathbf{Z}_2^N$. Denote by $w[l]$ ($1\le l\le N$) the vector in $\mathbf{Z}_2^N$ all components of which, except the $l$-th component, coincide with the corresponding components of the vector *w*, and denote by $S_w(f)$ the number of indices *l* for which $f(w[l]) \neq f(w)$. Then the *sensitivity* $S(f)$ of the function *f* is the maximum of $S_w(f)$ over all the vectors $w \in \mathbf{Z}_2^N$. For instance, if *f* is logical disjunction then $S(f) = N$, because in this case $f(0) = 0$ and $f(0[l]) = 1$ for any index *l*, whence $S_0(f) = N$. Analogously, if *f* is logical conjunction then $S(f) = N$ as well, because in this case $f(\mathbf{u}) = 1$ ($\mathbf{u}$ being the vector all the components of which are equal to one) and $f(\mathbf{u}[l]) = 0$ for any index *l*, so that $S_{\mathbf{u}}(f) = N$. There is a clear similarity between the vectors $w[l]$ for a given vector *w* and the entries of a logical matrix *A* that are neighbors of a given entry $a_{ij}$.

It may seem more natural to define the complexity of a logical matrix $A = (a_{ij})$ as the number $C_1(A)$ of pairs of unequal neighboring entries $(a_{ij}, a_{i+1,j})$ and $(a_{ij}, a_{i,j+1})$. However, very simple examples show that this definition of complexity is not quite adequate. For instance, consider two square logical matrices $A = (a_{ij})$ and $B = (b_{ij})$ of order 181, where $a_{ij} = 0$ for $j \leq 4$, $a_{ij} = 1$ for $j \geq 5$, $b_{ij} = 0$ for $j < i$ (i.e., below the main diagonal), and $b_{ij} = 1$ for $j \geq i$ (i.e., on and above the main diagonal). From an “intuitive” point of view, the complexity of these two matrices is more or less the same, and indeed, $C(A) = 362$ and $C(B) = 361$. At the same time, $C_1(A) = 181$ and $C_1(B) = 360$, i.e., the $C_1$-complexity of the matrix *B* is almost twice that of the matrix *A*.

Now let $G(E_{\text{col}}, b, r_{0k}, \kappa)$ be the complexity of the characteristic matrix of the two-dimensional region of recombination (7) for a given atom R and for given values of $E_{\text{col}}$, *b*, and $\zeta = \kappa r_{0k}$. A view of the two-dimensional recombination regions we have constructed shows that for fixed $E_{\text{col}}$ and *b*, the structure of the region in many cases does not depend very strongly on $r_{0k}$ and κ. In another work, we will examine the dependence of the quantity $G(E_{\text{col}}, b, r_{0k}, \kappa)$ averaged over ζ on the impact parameter *b* for a fixed collision energy $E_{\text{col}}$ (the so-called complexity function) and the connection of this dependence with the opacity functions of transcomplex recombination (7). Instead, here we will present images of five two-dimensional recombination regions with high complexities of the characteristic matrices.

In Table 2, for each atom R and for each of the nine values of the collision energy $E_{\text{col}}$ sampled, we present the maximal value $C_{\max}$ of the quantity $G(E_{\text{col}}, b, r_{0k}, \kappa)$ alongside the impact parameter $b$ and the parameter $\zeta = \kappa r_{0k}$ at which this maximum is attained. If the maximum $C_{\max}$ is attained at several pairs $(b, \zeta)$, then the pair with the minimal value of $b$ is pointed out. Finally, if for this value of the impact parameter $b$, the maximum $C_{\max}$ is attained at several values of the parameter $\zeta$, then we point out the value of $\zeta$ that is the earliest one in the sequence (5). Denote by $(b_A, \zeta_A)$ the pair $(b, \zeta)$ presented in Table 2 at which the maximum $C_{\max}$ of the quantity $G$ is attained. For each atom R and for each of the nine values of the collision energy $E_{\text{col}}$ sampled, we also present in Table 2 the minimum $C'$ of the quantity $G(E_{\text{col}}, b_A, r_{0k}, \kappa)$ over all the 14 values (5) of the parameter $\zeta$ alongside the value $\zeta = \zeta_Z$ at which this minimum is attained. If the minimum $C'$ is attained at several values of the parameter $\zeta$ then as $\zeta_Z$, we point out the value that is the earliest one in the sequence (5).

An analysis of Table 2 shows that, as was already mentioned above, two-dimensional regions of transcomplex recombination (7) with a strongly pronounced chaotic structure (i.e., with a high complexity of the characteristic matrix) are only encountered for low collision energies $E_{\text{col}}$. At $E_{\text{col}} \geq 0.3$ eV, the complexity $G(E_{\text{col}}, b, r_{0k}, \kappa)$ never exceeds 2589 for R = Kr and never exceeds 2175 for R = Xe. Analogously, at $E_{\text{col}} \geq 0.5$ eV, the complexity $G(E_{\text{col}}, b, r_{0k}, \kappa)$ for R = Hg never exceeds 2673. In Figs. 1–4, for R = Xe and Hg, we depict the two-dimensional recombination regions corresponding to the kinematic parameters $(E_{\text{col}}, b = b_A, \zeta = \zeta_A)$ with large values of $G = C_{\max}$, i.e., with $E_{\text{col}} = 0.1$ eV for R = Xe (Fig. 1) and with $0.1 \le E_{\text{col}} \le 0.3$ eV for R = Hg (Figs. 2–4).

Each of Figs. 1–4 as well as of Figs. 5–8 below is constructed as follows (cf. Figs. 5 and 6 in the paper [37]). For a given atom R and a given collection of the kinematic parameters $(E_{\text{col}}, b, \zeta)$, we consider the grid of all the points $(\theta, \varphi)$ with a step of $1°$ which we test successively. A point $(\theta, \varphi)$ corresponding to a non-recombinative trajectory for given R, $E_{\text{col}}$, $b$, and $\zeta$ (to be more precise, the small square centered at this point and with sides of $1°$) remains unpainted (white). A recombinative point $(\theta, \varphi)$ is painted one color or another depending on the value of the total internal energy $E_{\text{mol}}$ of the molecule CsBr formed. The color coding of the values of $E_{\text{mol}}$ is indicated on the legend strip at the top of each of Figs. 1–8. Note that in different Figs. 1–8, the same color gamut is used to represent different ranges of the total internal energy $E_{\text{mol}}$ of the CsBr product. The ranges of variation of $E_{\text{mol}}$ in Figs. 1–3, on the one hand, and in Figs. 4–8, on the other hand, differ drastically.

For a given atom R and a given collision energy $E_{\text{col}}$, consider the 14 two-dimensional regions of transcomplex recombination (7) for the same impact parameter $b = b_A$ and for various values of the parameter $\zeta$. The maximal complexity of the characteristic matrices of these regions is equal to $C_{\max}$ whereas the minimal complexity is equal to $C'$. Therefore, the difference between $C_{\max}$ and $C'$ in each line of Table 2 measures the dependence of the complexity $G(E_{\text{col}}, b, r_{0k}, \kappa)$ on $\zeta = \kappa r_{0k}$ (if $C' = C_{\max}$ then the complexity of the characteristic matrices of all the 14 two-dimensional regions is the same and $\zeta_Z = \zeta_A = r_{01}$). The largest difference $C_{\max} - C'$ equal to 8777 is observed for R = Kr and $E_{\text{col}} = 0.2$ eV ($C_{\max} = 10449$, $C' = 1672$). In this case, the discrepancy between the degrees of chaoticity of two-dimensional recombination regions differing only in the value of the parameter $\zeta$ is anomalously great. These two regions are shown in Figs. 5 and 6: the region for R = Kr with the kinematic parameters $E_{\mathrm{col}} = 0.2$ eV, $b = b_A$, $\zeta = \zeta_A$ in Fig. 5 and that for R = Kr with the kinematic parameters $E_{\mathrm{col}} = 0.2$ eV, $b = b_A$, $\zeta = \zeta_Z$ in Fig. 6. There is little in common between the regions in Figs. 5 and 6, although some structural elements (this primarily concerns the recombination-free "hole" with small θ angles elongated along the φ axis) are present in both the figures. What is also noteworthy is a small zone of very deep stabilization of the CsBr product (i.e., of small values of $E_{\mathrm{mol}}$) in the lower left corner of both the figures. The same zone is present in Fig. 4 as well as in Fig. 8 below. Somewhat paradoxically, to the right of this zone in Figs. 4–6 and 8, at a small distance, there is located a domain of non-recombinative points $(\theta, \varphi)$.

## 5. SURROUNDINGS INDEX OF A RECOMBINATIVE TRAJECTORY

A two-dimensional region of transcomplex recombination (7) can be quite "intricate" even in the case of low complexity $G(E_{\mathrm{col}}, b, r_{0k}, \kappa)$. For instance, Fig. 7 presents the two-dimensional recombination region for R = Kr, $E_{\mathrm{col}} = 0.5$ eV, $b = 17$ a.u., and $\zeta = r_{04} = 6.663$ a.u. with complexity $G = 1197$. This region can be tentatively called "an old man with a tobacco pipe". A similar appearance is exhibited by the two-dimensional regions for the same R, $E_{\mathrm{col}}$, and $b$ with $\zeta = r_{05} = 6.733$ a.u. $(G = 1215)$ and with $\zeta = r_{06} = 6.803$ a.u. $(G = 1251)$.

Here is another example. Figure 8 shows the two-dimensional recombination region for R = Kr, $E_{\mathrm{col}} = 0.2$ eV, $b = 19$ a.u., and $\zeta = -r_{01} = -6.452$ a.u. $(G = 870)$. For more visibility, this region is presented "in its entirety" rather than on the hemisphere corresponding to the angles $0 \leq \varphi \leq 180°$ (the characteristic matrix of the region is still constructed on the basis of the hemisphere $0 \leq \varphi \leq 180°$). Of course, the region in Fig. 8 is symmetric with respect to the "meridian" $\varphi = 180°$. The peculiarity of the region in Fig. 8 is the presence of four *isolated* points with coordinates $\theta = 11°$, $\varphi = 143°$ (in the “hole” with small angles θ and with angles φ smaller than $180°$), $\theta = 11°$, $\varphi = 217°$ (in the “hole” with small angles θ and with angles φ larger than $180°$), $\theta = 48°$, $\varphi = 170°$, and $\theta = 48°$, $\varphi = 190°$ (in the central “hole”). By an isolated point of a two-dimensional recombination region, we mean a recombinative point $(\theta, \varphi)$ for which all the eight points adjoining $(\theta, \varphi)$, i.e., the points $(\theta + \delta\theta, \varphi + \delta\varphi)$ with $\delta\theta = -1°; 0; 1°$, $\delta\varphi = -1°; 0; 1°$ ($\delta\theta$ and $\delta\varphi$ do not vanish simultaneously), are not recombinative. In this definition, we assume that $1° \le \theta \le 179°$. If a point $(\theta, \varphi)$ is one of the “poles” of the orientation sphere, i.e., if $\theta = 0$ or $\theta = 180°$, then for such a point, the definition of isolatedness should be modified in an obvious way.

Let us return to two-dimensional regions of recombination (7) for arbitrary R, $E_{\text{col}}$, *b*, and ζ on the hemisphere corresponding to the angles $0 \le \varphi \le 180°$. Consider any recombinative point $(\theta, \varphi)$ that coincides with none of the “poles” of the orientation sphere. There are eight points of the orientation sphere that *adjoin* this point: four neighboring points $(\theta - 1°, \varphi)$, $(\theta, \varphi - 1°)$, $(\theta + 1°, \varphi)$, $(\theta, \varphi + 1°)$ and four “adjacent” points $(\theta - 1°, \varphi - 1°)$, $(\theta + 1°, \varphi - 1°)$, $(\theta + 1°, \varphi + 1°)$, $(\theta - 1°, \varphi + 1°)$. Here for $\varphi = 0$, we take $1°$ in place of $\varphi - 1° = -1°$, while for $\varphi = 180°$, we take $179°$ in place of $\varphi + 1° = 181°$. Some of these eight points adjoining the recombinative point $(\theta, \varphi)$ may themselves be recombinative, some may not. Each recombinative point $(\theta, \varphi)$ with $1° \le \theta \le 179°$ and $0 \le \varphi \le 180°$ can be associated with the number *s* of adjoining points that are also recombinative. This number *s* (which can be called the *surroundings index* of the recombinative point in question and of the corresponding recombinative trajectory) ranges from 0 to 8. The value $s = 0$ indicates an isolated point of a two-dimensional recombination region.

For each atom R, for each of the nine values of the collision energy $E_{\text{col}}$ sampled, and for each of the nine values of the surroundings index *s* from 0 to 8, we have analyzed the corresponding assemblage of recombinative trajectories with $1° \le \theta \le 179°$, $0 \le \varphi \le 180°$ and for this assemblage, we have determined the number of trajectories $N$, the minimal ($E_{\text{mol},Z}$), maximal ($E_{\text{mol},A}$), and mean ($\langle E_{\text{mol}} \rangle$) total internal energy of the CsBr product, as well as the minimal ($T_Z$), maximal ($T_A$), and mean ($\langle T \rangle$) integration time of a trajectory. For R = Kr and for the first three values of $E_{\text{col}}$ as well as for the last value $E_{\text{col}} = 2.5$ eV, the results obtained are compiled in Table 3.

Of course, the time $T_A$ never exceeds the a priori utmost integration time $T_{\max} = 2.5 \times 10^7$ a.u. of a trajectory. On the other hand, as we already mentioned above, according to the equation (6) of the energy balance of a recombinative trajectory, the energy $E_{\text{mol},A}$ cannot exceed the smaller of the two quantities: $D = 4.39196$ eV and $E_{\text{col}} + D - D_c + \varepsilon_{0c} - V_{\text{st,min}}$ (see Table 1), i.e. (for R = Kr), it cannot exceed 4.26642 eV for $E_{\text{col}} = 0.1$ eV, it cannot exceed 4.36642 eV for $E_{\text{col}} = 0.2$ eV, and it cannot exceed 4.39196 eV for $E_{\text{col}} \ge 0.3$ eV. This is consistent with the data in Table 3. In the work [36], we presented the energy balance equation of a recombinative trajectory where $V_{\text{i}}$ was used in place of $V_{\text{st}}$ (see the equation (4) in the paper [36]). Such an equation is not quite correct. If it were true, then the total internal energy $E_{\text{mol}}$ of the CsBr molecule could not exceed $\min\{D; E_{\text{col}} + D - D_c + \varepsilon_{0c} - V_{\text{i}}\}$ (see Table 1) for any recombinative trajectory. In particular, the energy $E_{\text{mol}}$ could not exceed 4.26528 eV for $E_{\text{col}} = 0.1$ eV and could not exceed 4.36528 eV for $E_{\text{col}} = 0.2$ eV. At the same time, some energies $E_{\text{mol},A}$ for $E_{\text{col}} = 0.1$ and 0.2 eV in Table 3 are larger than those values.

One sees in Table 3 that for R = Kr, at each value of the collision energy $E_{\text{col}}$, the total internal energy $E_{\text{mol}}$ of the CsBr product and the integration time $T$ of a recombinative trajectory generally decrease as the surroundings index $s$ increases. This holds for both the mean and minimal values of $E_{\text{mol}}$ and $T$. The most interesting situation, in our opinion, occurs for $0.3 \le E_{\text{col}} \le 2.5$ eV (Table 3 only lists the data for two extreme values of $E_{\text{col}}$ within this interval). For these collision energies, as one passes from $s=7$ (when one of the points of the orientation sphere adjoining the given point $(\theta,\varphi)$ is not recombinative) to $s=8$ (when the given point $(\theta,\varphi)$ lies "strictly inside" the two-dimensional recombination region), the minimal total internal energy $E_{\text{mol},Z}$ of the CsBr molecule and the mean integration time $\langle T\rangle$ of a trajectory are reduced very sharply, whereas the mean total internal energy $\langle E_{\text{mol}}\rangle$ of the CsBr product and the minimal integration time $T_Z$ of a trajectory also decrease noticeably.

For R = Xe and Hg, the behavior of the dependences of the quantities $E_{\text{mol},Z}$, $\langle E_{\text{mol}}\rangle$, $E_{\text{mol},A}$, $T_Z$, $\langle T\rangle$, and $T_A$ on $E_{\text{col}}$ and *s* is almost the same as for R = Kr. For each value of $E_{\text{col}}$, the energy $E_{\text{mol}}$ and the time *T* generally decrease as the surroundings index *s* increases, and this holds for both the mean and minimal values. For $0.3 \le E_{\text{col}} \le 2.5$ eV in the case where R = Xe and for $0.5 \le E_{\text{col}} \le 2.5$ eV in the case where R = Hg, "crucial" changes in $E_{\text{mol}}$ and *T* occur as one passes from $s=7$ to $s=8$: the energy $E_{\text{mol},Z}$ drops sharply to very low values, the time $\langle T\rangle$ is strongly reduced as well, whereas the decrease in the quantities $\langle E_{\text{mol}}\rangle$ and $T_Z$, although distinct, is less sharp.

Thus, for each atom R, there exists a noticeable correlation among the total internal energy $E_{\text{mol}}$ of the CsBr molecule, the integration time *T* of a recombinative trajectory, and the surroundings index of that trajectory, i.e., the number *s* of recombinative points adjoining the point $(\theta,\varphi)$ corresponding to the trajectory in question. The larger *s*, the shorter the trajectory on the whole (i.e., the smaller *T*) and the deeper the stabilization of the recombination product (i.e., the smaller $E_{\text{mol}}$).

Of course, the energy $E_{\text{mol},A}$ for R = Xe and Hg, as for R = Kr, never exceeds $\min\{D; E_{\text{col}} + D - D_c + \varepsilon_{0c} - V_{\text{st,min}}\}$. For R = Xe, the latter quantity is equal to 4.27849 eV for $E_{\text{col}} = 0.1$ eV, is equal to 4.37849 eV for $E_{\text{col}} = 0.2$ eV, and is equal to $D = 4.39196$ eV for $E_{\text{col}} \ge 0.3$ eV (see Table 1). For R = Hg, the quantity $\min\{D; E_{\rm col} + D - D_c + \varepsilon_{0c} - V_{\rm st,min}\}$ is equal to 4.18428, 4.28428, 4.38428 eV and is equal to $D$ for $E_{\rm col} = 0.1$, 0.2, 0.3 eV and for $E_{\rm col} \geq 0.5$ eV, respectively. At the same time, for R = Xe at $E_{\rm col} = 0.1$ eV, the energy $E_{{\rm mol},A}$ sometimes exceeds $E_{\rm col} + D - D_c + \varepsilon_{0c} - V_{\rm i} = 4.27683$ eV (see Table 1). For R = Hg at $E_{\rm col} = 0.1$ and 0.2 eV, the energy $E_{{\rm mol},A}$ also sometimes exceeds $E_{\rm col} + D - D_c + \varepsilon_{0c} - V_{\rm i}$ (the latter quantity is equal to 4.18247 and 4.28247 eV at $E_{\rm col} = 0.1$ and 0.2 eV, respectively). For unclear reasons, the energy $E_{{\rm mol},A}$ in our calculations did not exceed 4.367 eV for any R, $E_{\rm col}$, and *s*.

## 6. DISTRIBUTIONS OF THE ORIENTATION ANGLES

The complicated nature of the dynamics of transcomplex recombination (7) is illustrated by the dependences of the mean values $\langle\theta\rangle$ and $\langle\varphi\rangle$ of the orientation angles $\theta$ and $\varphi$ for recombinative trajectories on the impact parameter *b* for a given atom R and a given collision energy $E_{\rm col}$. The averaging is carried out over the entire set of recombinative trajectories with $0 \leq \varphi \leq 180°$ for the given values of $E_{\rm col}$ and *b*. The "bizarre" dependences $\langle\theta\rangle(b) = \langle\theta\rangle(E_{\rm col}, b)$ and $\langle\varphi\rangle(b) = \langle\varphi\rangle(E_{\rm col}, b)$ for R = Kr are shown in Figs. 9 and 10, whereas for R = Hg they are shown in Figs. 11 and 12. The dependences $\langle\theta\rangle(b)$ and $\langle\varphi\rangle(b)$ for R = Xe differ slightly from the corresponding dependences for R = Kr. The behavior of the dependences $\langle\theta\rangle(b)$ and $\langle\varphi\rangle(b)$ apparently cannot be explained without an analysis of the detailed dynamics of transcomplex recombination (7), which will be the subject of subsequent publications. Note that since for $b = 0$ trajectories that differ only in the value of the $\varphi$ angle are taken to each other by rotations about the *x* axis, $\langle\varphi\rangle(0) = 90°$ for any R and $E_{\rm col}$.

To clarify the procedure for constructing the graphs of the functions $\langle\theta\rangle(b)$ and $\langle\varphi\rangle(b)$ in Figs. 9–12, consider the case where R = Kr and $E_{\rm col} = 0.1$ eV. In this case, the maximal value of the impact parameter $b$ at which we observed recombinative trajectories was equal to 110 a.u. Moreover, at $b = 110$ a.u., only four recombinative trajectories with $\varphi \le 180°$ were recorded. The initial conditions of these trajectories and the total internal energy $E_{mol}$ of the CsBr product are

$$(\zeta = r_{02} = 6.522 \text{ a.u.}, \ \theta = 3°, \ \varphi = 169°, \ E_{mol} = 4.143 \text{ eV}),$$

$$(\zeta = r_{04} = 6.663 \text{ a.u.}, \ \theta = 15°, \ \varphi = 123°, \ E_{mol} = 4.08 \text{ eV}),$$

$$(\zeta = r_{05} = 6.733 \text{ a.u.}, \ \theta = 4°, \ \varphi = 167°, \ E_{mol} = 4.066 \text{ eV}),$$

$$(\zeta = -r_{07} = -6.873 \text{ a.u.}, \ \theta = 7°, \ \varphi = 49°, \ E_{mol} = 4.097 \text{ eV}).$$

Calculating the arithmetic means of these values of the orientation angles θ and φ, we get the following numbers for R = Kr:

$$\langle\theta\rangle \text{(0.1 eV, 110 a.u.)} = \tfrac{1}{4}(3 + 15 + 4 + 7) = \tfrac{29}{4} = 7.25°$$

(this is how the extreme right point on line *1* in Fig. 9*a* is obtained),

$$\langle\varphi\rangle \text{(0.1 eV, 110 a.u.)} = \tfrac{1}{4}(169 + 123 + 167 + 49) = \tfrac{508}{4} = 127°$$

(this is how the extreme right point on line *1* in Fig. 10*a* is obtained).

As additional characteristics of transcomplex recombination (7) for a given atom R, one can consider the functions $L(\theta, \varphi)$ and $\Xi(\theta, \varphi)$ on the orientation sphere (for $0 \le \varphi \le 180°$). The quantity $L(\theta, \varphi)$ is the number of collections $(E_{col}, b, \zeta)$ in our calculations for which the trajectory with the given angles θ and φ for the given atom R is recombinative. One can interpret the value $L(\theta, \varphi)$ as a measure of how “favorable” one or another pair of the angles $(\theta, \varphi)$ is for recombination (7). In the paper [37], we dealt with an analogous function $L(\Theta, \Phi)$ for direct three-body recombination (8). The quantity $\Xi(\theta, \varphi)$ is defined as follows. Let $(\theta, \varphi)$ be a certain pair of the orientation angles measured in degrees. For the given atom R, take some values of the kinematic parameters $E_{col}$, $b$, $\zeta = \kappa r_{0k}$ and consider the characteristic matrix $A = (a_{ij})$ of the corresponding two-dimensional recombination region and the entry $a_{\varphi+1, \theta+1}$ of $A$. Then $\Xi(\theta, \varphi)$ is the number of collections $(E_{col}, b, \zeta)$ in our calculations for which the entry $a_{\varphi+1,\theta+1}$ of the characteristic matrix possesses at least one neighboring entry not equal to $a_{\varphi+1,\theta+1}$ (i.e., a discordant neighboring entry). One can regard the quantity $\Xi(\theta,\varphi)$ as a measure of how effectively one or another pair of the angles $(\theta,\varphi)$ "contributes" to the complexity $C(A)$ of the characteristic matrices $A$ of two-dimensional recombination regions.

Since for $\theta = 0$ and $\theta = 180°$, the initial conditions of a trajectory are independent of the $\varphi$ angle, the values of $L(0,\varphi)$ and $L(180°,\varphi)$ are independent of the $\varphi$ angle as well. For R = Kr, there hold the identities $L(0,\varphi) \equiv 3302$ and $L(180°,\varphi) \equiv 4116$, for R = Xe, the identities $L(0,\varphi) \equiv 3798$ and $L(180°,\varphi) \equiv 5080$, and for R = Hg, the identities $L(0,\varphi) \equiv 3691$ and $L(180°,\varphi) \equiv 4808$. Moreover, for each atom R, the sum of the values $L(\theta,\varphi)$ over all the angles $\theta$ and $\varphi$ from 0 to $180°$ is equal to the total number of recombinative trajectories which we recorded for that atom R (127423119 for R = Kr, 151532118 for R = Xe, and 148538851 for R = Hg).

The sum of the values $\Xi(\theta,\varphi)$ over all the angles $\theta$ and $\varphi$ from 0 to $180°$ for each atom R is equal to the sum of the complexities of the characteristic matrices of all the two-dimensional recombination regions for that atom R. This sum is equal to 5393814 for R = Kr, to 5102685 for R = Xe, and to 5941465 for R = Hg.

In Table 4, for all the three atoms R, we present the minimal and maximal values of the functions $L(\theta,\varphi)$ and $\Xi(\theta,\varphi)$ as well as the pairs of the angles $(\theta,\varphi)$ at which these extrema are attained (in all the cases, the extremum is attained at a single point, provided that one confines oneself with the $\varphi$ angles no greater than $180°$). As examples, Figs. 13 and 14 depict the two-dimensional graphs of the functions $L(\theta,\varphi)$ and $\Xi(\theta,\varphi)$ for R = Hg. The graphs of these functions for R = Kr and Xe look similar. The most characteristic features of the graphs of the functions $L(\theta,\varphi)$ and $\Xi(\theta,\varphi)$ are the very sharp minimum of the function $L(\theta,\varphi)$ at a point $(\theta = \theta_m, \varphi = 0)$ where $\theta_m$ is almost independent of R and ranges in the interval $30° \le \theta_m \le 36°$ and the very sharp maximum of the function $\Xi(\theta,\varphi)$ at a close point $(\theta = \theta_M, \varphi = 0)$ where $\theta_M$ is also almost independent of R and ranges in the interval $24° \le \theta_M \le 31°$ (see Table 4). The difference $\theta_m - \theta_M$ is equal to $4°$, $5°$, and $6°$ for R = Kr, Xe, and Hg, respectively. There is practically no doubt that it is impossible to explain the presence of such a minimum of the function $L(\theta,\varphi)$ and of such a maximum of the function $\Xi(\theta,\varphi)$ without analyzing the detailed dynamics of transcomplex recombination (7). Note that the points $(\theta = \theta_m, \varphi = 0)$ and $(\theta = \theta_M, \varphi = 0)$ lie in the vicinity of the zone of deep stabilization of the CsBr product in Figs. 4–6 and 8 for R = Kr and Hg (see the end of Section 4).

## 7. CONCLUSIONS

A chaotic structure of many two-dimensional regions of transcomplex recombination (7) on the orientation sphere with the coordinates $(\theta,\varphi)$ shows that the dynamics of transcomplex recombination (2) is much more complicated than the dynamics of direct three-body recombination (1). In this paper, we presented images of five two-dimensional regions of recombination (7) with a high degree of chaoticity (Figs. 1–5). These images demonstrate that the chaotic two-dimensional regions of transcomplex recombination (7) are highly diverse. On the other hand, the analysis of various characteristics of recombinative trajectories allows one to identify a number of patterns hidden “behind the facade” of chaos. For instance, the integration time $T$ of a recombinative trajectory and the total internal energy $E_{\text{mol}}$ of the product molecule CsBr are closely connected with the surroundings index $s$ of the trajectory. On the whole, the larger $s$, the smaller $T$ and $E_{\text{mol}}$. Moreover, for not very low collision energies $E_{\text{col}}$, the most noticeable changes in $T$ and $E_{\text{mol}}$ occur as one passes from $s = 7$ to the maximal value $s = 8$. Another manifestation of the complicated nature of the dynamics of transcomplex recombination (7) is the peculiar behavior of the functions $\langle\theta\rangle(b)$, $\langle\varphi\rangle(b)$, $L(\theta,\varphi)$, and $\Xi(\theta,\varphi)$ which were discussed in Section 6.

It is not clear how the structure of the two-dimensional regions of transcomplex recombination (7) and various dynamical characteristics of recombinative trajectories can change if one uses another step of the grid over which one successively tests the values of the orientation angles θ and φ (for instance, if one makes this step smaller than $1^\circ$ along both the angles) or if one utilizes another criterion of the occurrence of recombination (for example, if one sets the utmost integration time $T_{\max}$ of a trajectory to be larger than $2.5\times10^7$ a.u.). This question will be the subject of further research, the results of which will likely be far from trivial. For instance, it is obvious that while refining the angle grid, the surroundings index of a recombinative trajectory may become completely different, as the collection of the adjoining points of the orientation sphere will change. Note that the structure of the two-dimensional regions of direct three-body recombination (8) on the sphere of unit radius with the coordinates $(\Theta,\Phi)$ is apparently not very sensitive to such details of the calculation procedure [37].

As was already mentioned in Section 4, in another work we will examine the connection between the opacity functions $p(b)=p(E_{\text{col}},b)$ of transcomplex recombination (7) and the complexity functions $q(b)=q(E_{\text{col}},b)$. The value $q(b)$ is the result of averaging of the complexity $G(E_{\text{col}},b,r_{0k},\kappa)$ of the characteristic matrix of a two-dimensional recombination region over $r_{0k}$ and κ. One may call the functions $p(b)$ and $q(b)$ as well as the functions $\langle\theta\rangle(b)$ and $\langle\varphi\rangle(b)$ (the mean values of the orientation angles for recombinative trajectories) *intracomplex* functions because to calculate them, one fixes the collision energy $E_{\text{col}}$ and the impact parameter $b$ and then deals with the entire range of values of the kinematic parameters θ, φ, and ζ which describe the $RCs^+$ complex at the beginning of the trajectory.

Of course, what also deserves attention is the patterns of the distributions of the vibrational energy $E_{\mathrm{vib}}$ and of the rotational energy $E_{\mathrm{rot}}$ of the CsBr molecule as well as of the energy $E_{\mathrm{fly}}$ of the flying apart of the products CsBr and R. Finally, the question is also of interest what is the structure of the two-dimensional regions of transcomplex recombination (2) for other systems $RCs^+ + X^-$, for example, for the system $ArCs^+ + Br^-$ (cf. the work [40]) or for the systems $RCs^+ + X^-$ with R = Ar, Xe and $X^- = F^-$, $I^-$ which were considered in the papers [34, 36]. We plan to devote some further research to these topics.

## FUNDING

This study was carried out in the framework of State Assignment of the Ministry of Science and Higher Education of the Russian Federation (project No. 125012200611-5).

## CONFLICT OF INTEREST

The authors of this work declare that they have no conflicts of interest.

**Table 1.** Energy characteristics of the transcomplex recombination reactions (7) in eV

| R | $D_c$ | $\varepsilon_{0c}$ | $V_{\mathrm{st,min}}$ | $D - D_c + \varepsilon_{0c} - V_{\mathrm{st,min}}$ | $D - D_c + \varepsilon_{0c} - V_{\mathrm{i}}$ |
|---|---|---|---|---|---|
| Kr | 0.121031 | 0.00318916 | 0.107701 | 4.16642 | 4.16528 |
| Xe | 0.108505 | 0.00221724 | 0.107182 | 4.17849 | 4.17683 |
| Hg | 0.204004 | 0.00335857 | 0.107038 | 4.08428 | 4.08247 |

**Table 2.** The maximal complexity $C_{\max}$ of the characteristic matrix of a two-dimensional region of recombination (7) for the given atom R and the given collision energy $E_{\text{col}}$, the values $b_A$ and $\zeta_A$ of the kinematic parameters $b$ and $\zeta$ at which this maximum is attained, the minimal complexity $C'$ of the characteristic matrix for $b = b_A$, and the value $\zeta_Z$ of the parameter $\zeta$ at which this minimum is attained

| $E_{\text{col}}$, eV | $C_{\max}$ | $b_A$, a.u. | $\zeta_A$, a.u. | $C'$ | $\zeta_Z$, a.u. |
|---|---|---|---|---|---|
| R = Kr | | | | | |
| 0.1 | 17037 | 97 | $r_{01} = 6.452$ | 15981 | $-r_{03} = -6.592$ |
| 0.2 | 10449 | 21 | $r_{03} = 6.592$ | 1672 | $-r_{05} = -6.733$ |
| 0.3 | 2589 | 17 | $-r_{02} = -6.522$ | 1824 | $r_{07} = 6.873$ |
| 0.5 | 1553 | 9 | $-r_{05} = -6.733$ | 1374 | $-r_{07} = -6.873$ |
| 0.75 | 1253 | 14 | $-r_{02} = -6.522$ | 514 | $r_{06} = 6.803$ |
| 1 | 1086 | 0 | $r_{01} = 6.452$ | 1086 | $r_{01} = 6.452$ |
| 1.5 | 1086 | 0 | $r_{01} = 6.452$ | 1086 | $r_{01} = 6.452$ |
| 2 | 1086 | 0 | $r_{01} = 6.452$ | 1086 | $r_{01} = 6.452$ |
| 2.5 | 1086 | 0 | $r_{01} = 6.452$ | 1086 | $r_{01} = 6.452$ |
| R = Xe | | | | | |
| 0.1 | 16083 | 105 | $r_{06} = 7.735$ | 12634 | $-r_{01} = -7.364$ |
| 0.2 | 3793 | 22 | $-r_{03} = -7.513$ | 1619 | $-r_{04} = -7.587$ |
| 0.3 | 2175 | 18 | $r_{01} = 7.364$ | 1321 | $-r_{04} = -7.587$ |
| 0.5 | 1527 | 9 | $r_{07} = 7.81$ | 1331 | $-r_{06} = -7.735$ |
| 0.75 | 1421 | 14 | $-r_{07} = -7.81$ | 884 | $-r_{02} = -7.438$ |
| 1 | 1086 | 0 | $r_{01} = 7.364$ | 1086 | $r_{01} = 7.364$ |
| 1.5 | 1086 | 0 | $r_{01} = 7.364$ | 1086 | $r_{01} = 7.364$ |
| 2 | 1094 | 1 | $r_{04} = 7.587$ | 1086 | $r_{01} = 7.364$ |
| 2.5 | 1176 | 1 | $-r_{04} = -7.587$ | 1086 | $r_{03} = 7.513$ |
| R = Hg | | | | | |
| 0.1 | 16895 | 97 | $-r_{02} = -6.746$ | 9012 | $r_{05} = 6.909$ |
| 0.2 | 11448 | 71 | $-r_{06} = -6.964$ | 8408 | $-r_{01} = -6.692$ |
| 0.3 | 5360 | 17 | $r_{02} = 6.746$ | 1478 | $r_{04} = 6.855$ |
| 0.5 | 2673 | 14 | $r_{06} = 6.964$ | 2260 | $-r_{07} = -7.018$ |
| 0.75 | 1810 | 0 | $r_{01} = 6.692$ | 1810 | $r_{01} = 6.692$ |
| 1 | 1662 | 8 | $-r_{01} = -6.692$ | 929 | $r_{07} = 7.018$ |
| 1.5 | 1086 | 0 | $r_{01} = 6.692$ | 1086 | $r_{01} = 6.692$ |
| 2 | 1294 | 1 | $r_{07} = 7.018$ | 1086 | $r_{01} = 6.692$ |
| 2.5 | 1418 | 1 | $r_{03} = 6.801$ | 1096 | $-r_{06} = -6.964$ |

**Table 3.** The statistics of the recombinative trajectories with R = Kr (and with $1° \le \theta \le 179°$, $0 \le \varphi \le 180°$) for $E_{\rm col} = 0.1$, 0.2, 0.3, and 2.5 eV and for various values of the surroundings index *s*. The energies $E_{{\rm mol},Z}$, $\langle E_{\rm mol} \rangle$, and $E_{{\rm mol},A}$ are given in eV, whereas the integration times $T_Z$, $\langle T \rangle$, and $T_A$ are given in a.u.

| $s$ | $N$ | $E_{{\rm mol},Z}$ | $\langle E_{\rm mol} \rangle$ | $E_{{\rm mol},A}$ | $T_Z$ | $\langle T \rangle$ | $T_A$ |
|---|---|---|---|---|---|---|---|
| $E_{\rm col} = 0.1$ eV | | | | | | | |
| 0 | 53704 | 3.813 | 4.017 | 4.258 | 3863100 | 15392213 | 25000000 |
| 1 | 88848 | 0.933 | 3.997 | 4.260 | 2087000 | 12724163 | 24999000 |
| 2 | 158353 | 0.127 | 3.970 | 4.258 | 1622800 | 9990999 | 24999950 |
| 3 | 142682 | 0.150 | 3.968 | 4.264 | 1730100 | 9392563 | 24999950 |
| 4 | 204708 | 0.086 | 3.971 | 4.264 | 1529550 | 8382490 | 24999200 |
| 5 | 288399 | 0.047 | 3.946 | 4.266 | 850350 | 6970046 | 25000000 |
| 6 | 199946 | 0.041 | 3.958 | 4.266 | 1191200 | 6898415 | 24999950 |
| 7 | 302359 | 0.032 | 3.907 | 4.266 | 1039600 | 6384235 | 24998600 |
| 8 | 42070919 | 0.015 | 3.691 | 4.266 | 788550 | 1553411 | 24997100 |
| $E_{\rm col} = 0.2$ eV | | | | | | | |
| 0 | 436 | 0.564 | 3.993 | 4.365 | 12426050 | 22921960 | 24999700 |
| 1 | 3060 | 0.148 | 4.005 | 4.364 | 2531700 | 22257494 | 24998800 |
| 2 | 11891 | 0.099 | 4.004 | 4.365 | 1651200 | 20913927 | 24999800 |
| 3 | 26358 | 0.138 | 3.998 | 4.365 | 1644450 | 20136633 | 24999900 |
| 4 | 55663 | 0.021 | 4.032 | 4.366 | 1278450 | 16912099 | 24999750 |
| 5 | 125125 | 0.043 | 4.073 | 4.366 | 816350 | 11468615 | 24999700 |
| 6 | 97327 | 0.068 | 4.061 | 4.366 | 840500 | 12474260 | 24999450 |
| 7 | 134589 | 0.032 | 4.131 | 4.366 | 825600 | 9738163 | 24997150 |
| 8 | 29177674 | 0.135 | 3.769 | 4.366 | 658950 | 1413974 | 24998150 |
| $E_{\rm col} = 0.3$ eV | | | | | | | |
| 0 | 1190 | 4.302 | 4.351 | 4.367 | 3916100 | 14047970 | 24983100 |
| 1 | 1863 | 4.259 | 4.355 | 4.367 | 2400450 | 16348982 | 24999300 |
| 2 | 5093 | 4.180 | 4.351 | 4.367 | 1403850 | 14787257 | 24999250 |
| 3 | 2798 | 4.194 | 4.351 | 4.367 | 1508900 | 15161242 | 24999600 |
| 4 | 51777 | 4.172 | 4.357 | 4.367 | 1365550 | 18662609 | 24999950 |
| 5 | 233338 | 3.366 | 4.327 | 4.367 | 739100 | 11558345 | 25000000 |
| 6 | 62652 | 3.363 | 4.318 | 4.367 | 738900 | 10158315 | 24994100 |
| 7 | 83994 | 3.356 | 4.311 | 4.367 | 738650 | 9511866 | 24999550 |
| 8 | 22719042 | 0.245 | 3.800 | 4.367 | 588150 | 1598339 | 24964250 |
| $E_{\rm col} = 2.5$ eV | | | | | | | |
| 3 | 43 | 4.362 | 4.366 | 4.367 | 18088950 | 23091669 | 24988550 |
| 4 | 8411 | 4.232 | 4.351 | 4.367 | 1635200 | 13869826 | 24997950 |
| 5 | 31017 | 3.697 | 4.271 | 4.367 | 356800 | 4909819 | 24988800 |
| 6 | 8140 | 3.698 | 4.264 | 4.365 | 357050 | 3194106 | 21616000 |
| 7 | 11445 | 3.690 | 4.234 | 4.363 | 356900 | 2282890 | 19014200 |
| 8 | 1046132 | 0.120 | 2.778 | 4.356 | 298450 | 415976 | 14281550 |

For $E_{\rm col} = 2.5$ eV, we have not encountered recombinative trajectories with $0 \le s \le 2$ ($N = 0$).

**Table 4.** The minimal values $L_{\min}, \Xi_{\min}$ and the maximal values $L_{\max}, \Xi_{\max}$ of the functions $L(\theta, \varphi)$ and $\Xi(\theta, \varphi)$ alongside the angle pairs $(\theta, \varphi)$ at which these extrema are attained

| R | $L_{\min}$ | $L_{\max}$ | $\Xi_{\min}$ | $\Xi_{\max}$ |
|---|---|---|---|---|
| Kr | 2862, (32°,0°) | 4278, (124°,166°) | 7, (180°,88°) | 817, (28°,0°) |
| Xe | 2947, (36°,0°) | 5102, (172°,19°) | 3, (180°,90°) | 1016, (31°,0°) |
| Hg | 3041, (30°,0°) | 4894, (125°,156°) | 9, (180°,90°) | 1102, (24°,0°) |

Figure captions

**Fig. 1.** The two-dimensional region of transcomplex recombination $XeCs^{+} + Br^{-}$ on the orientation sphere $(\theta, \varphi)$ at $E_{col} = 0.1$ eV, $b = 105$ a.u., and $\zeta = 7.735$ a.u. The complexity $G$ of the characteristic matrix of this region is equal to 16083.

**Fig. 2.** The two-dimensional region of transcomplex recombination $HgCs^{+} + Br^{-}$ on the orientation sphere $(\theta, \varphi)$ at $E_{col} = 0.1$ eV, $b = 97$ a.u., and $\zeta = -6.746$ a.u. The complexity $G$ of the characteristic matrix of this region is equal to 16895.

**Fig. 3.** The two-dimensional region of transcomplex recombination $HgCs^{+} + Br^{-}$ on the orientation sphere $(\theta, \varphi)$ at $E_{col} = 0.2$ eV, $b = 71$ a.u., and $\zeta = -6.964$ a.u. The complexity $G$ of the characteristic matrix of this region is equal to 11448.

**Fig. 4.** The two-dimensional region of transcomplex recombination $HgCs^{+} + Br^{-}$ on the orientation sphere $(\theta, \varphi)$ at $E_{col} = 0.3$ eV, $b = 17$ a.u., and $\zeta = 6.746$ a.u. The complexity $G$ of the characteristic matrix of this region is equal to 5360.

**Fig. 5.** The two-dimensional region of transcomplex recombination $KrCs^{+} + Br^{-}$ on the orientation sphere $(\theta, \varphi)$ at $E_{col} = 0.2$ eV, $b = 21$ a.u., and $\zeta = 6.592$ a.u. The complexity $G$ of the characteristic matrix of this region is equal to 10449.

**Fig. 6.** The two-dimensional region of transcomplex recombination $KrCs^{+} + Br^{-}$ on the orientation sphere $(\theta, \varphi)$ at $E_{col} = 0.2$ eV, $b = 21$ a.u., and $\zeta = -6.733$ a.u. The complexity $G$ of the characteristic matrix of this region is equal to 1672.

**Fig. 7.** The two-dimensional region of transcomplex recombination $KrCs^+ + Br^-$ on the orientation sphere $(\theta, \varphi)$ at $E_{col} = 0.5$ eV, $b = 17$ a.u., and $\zeta = 6.663$ a.u. The complexity $G$ of the characteristic matrix of this region is equal to 1197.

**Fig. 8.** The two-dimensional region of transcomplex recombination $KrCs^+ + Br^-$ on the orientation sphere $(\theta, \varphi)$ for $0 \le \varphi < 360°$ at $E_{col} = 0.2$ eV, $b = 19$ a.u., and $\zeta = -6.452$ a.u. The complexity $G$ of the characteristic matrix of this region is equal to 870.

**Fig. 9.** The dependences $\langle\theta\rangle(b)$ for transcomplex recombination $KrCs^+ + Br^-$. Panel *a*: lines *1*, *2*, *3*, and *4* correspond to the collision energies $E_{col} = 0.1$, 0.2, 0.3, and 0.5 eV, respectively. Panel *b*: lines *1*, *2*, *3*, *4*, and *5* correspond to the collision energies $E_{col} = 0.75$, 1, 1.5, 2, and 2.5 eV, respectively.

**Fig. 10.** The dependences $\langle\varphi\rangle(b)$ for transcomplex recombination $KrCs^+ + Br^-$. The lines in panels *a* and *b* have the same meaning as those in Fig. 9.

**Fig. 11.** The dependences $\langle\theta\rangle(b)$ for transcomplex recombination $HgCs^+ + Br^-$. The lines in panels *a* and *b* have the same meaning as those in Fig. 9.

**Fig. 12.** The dependences $\langle\varphi\rangle(b)$ for transcomplex recombination $HgCs^+ + Br^-$. The lines in panels *a* and *b* have the same meaning as those in Fig. 9.

**Fig. 13.** The graph of the function $L(\theta, \varphi)$ for transcomplex recombination $HgCs^+ + Br^-$.

**Fig. 14.** The graph of the function $\Xi(\theta, \varphi)$ for transcomplex recombination $HgCs^+ + Br^-$.

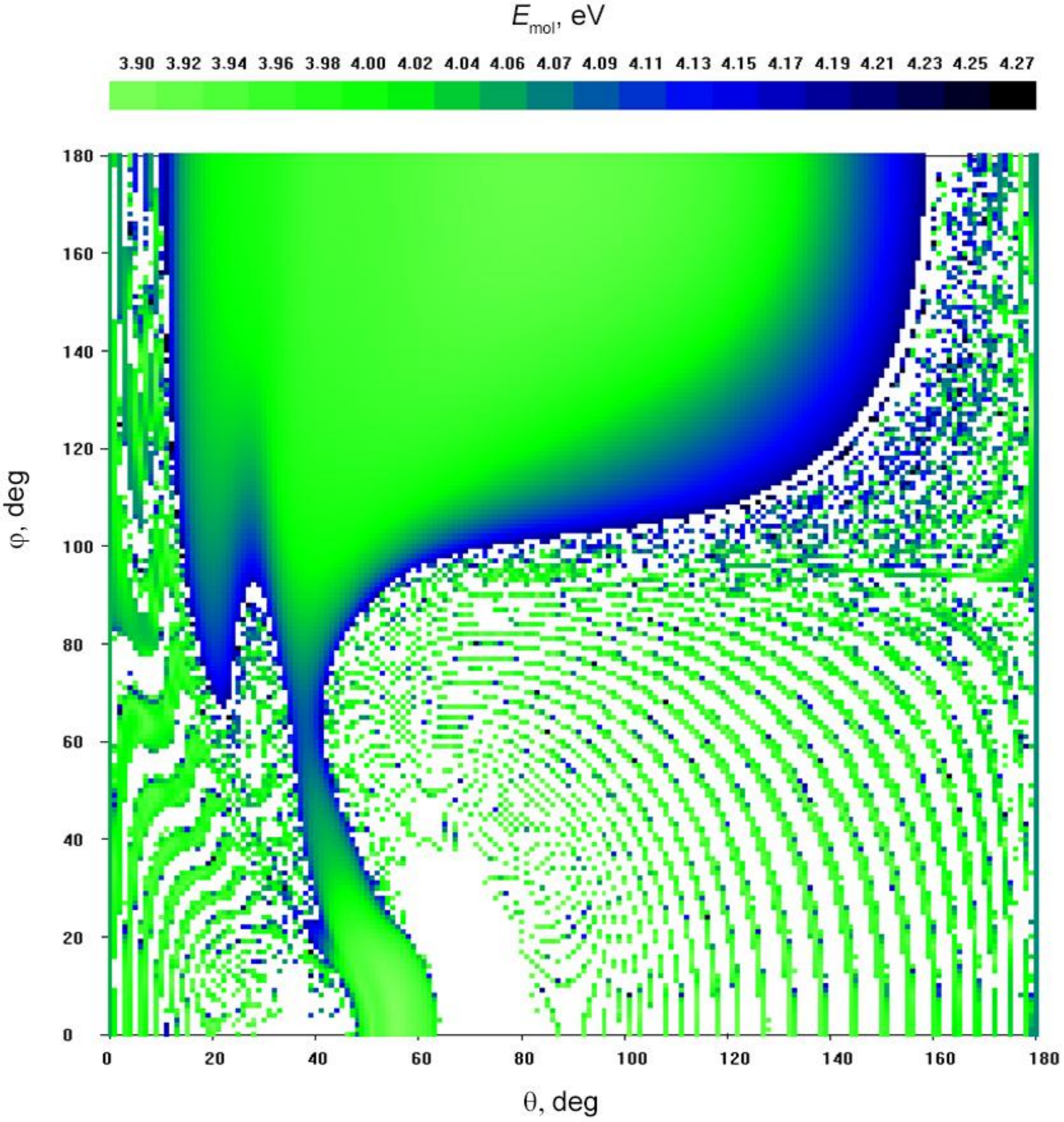

**Fig. 1** for the paper by *V. M. Akimov et al.* “The existence regions for transcomplex recombination of heavy ions. Two-dimensional recombination regions and surroundings indices”

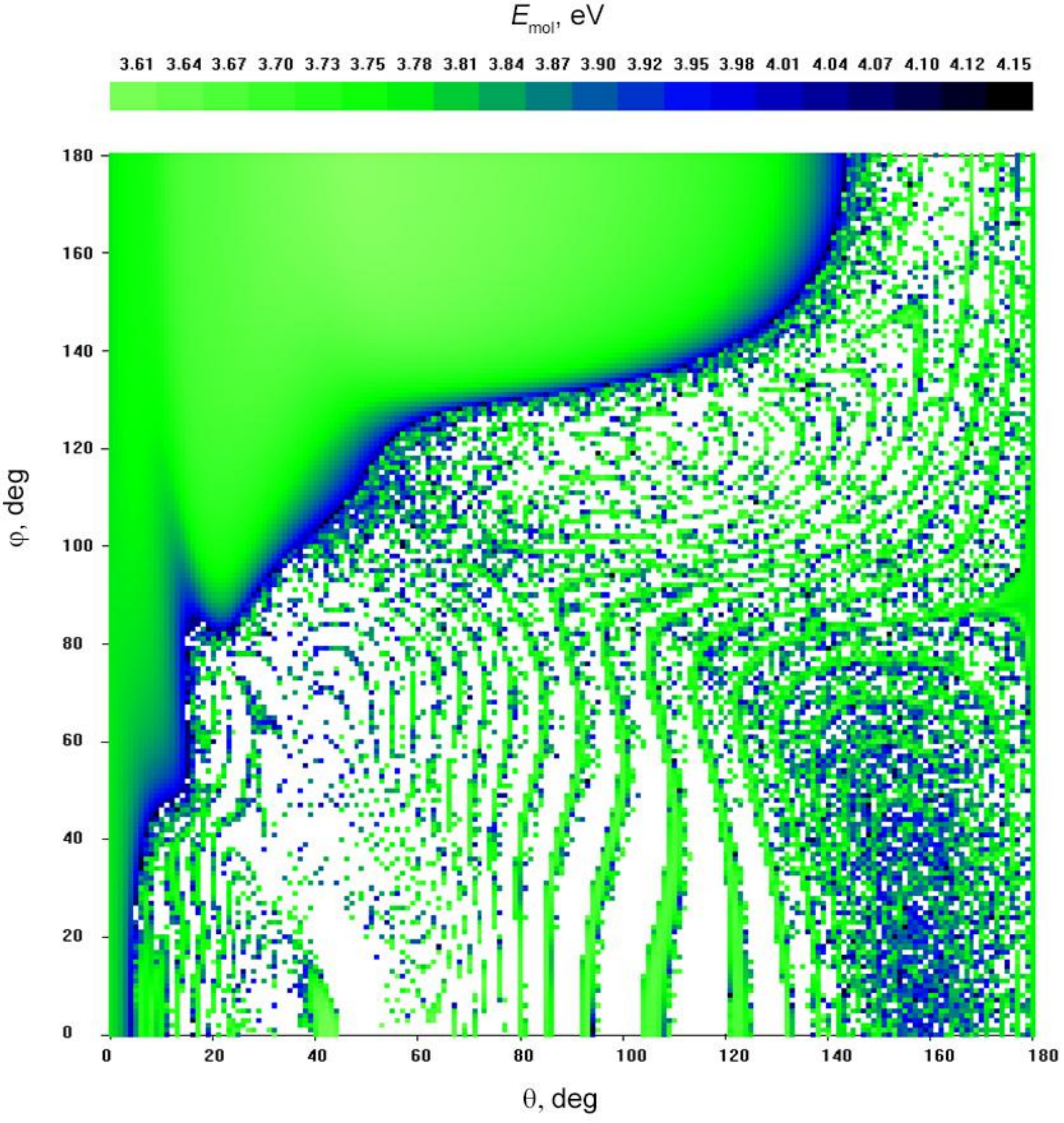

**Fig. 2** for the paper by *V. M. Akimov et al.* “The existence regions for transcomplex recombination of heavy ions. Two-dimensional recombination regions and surroundings indices”

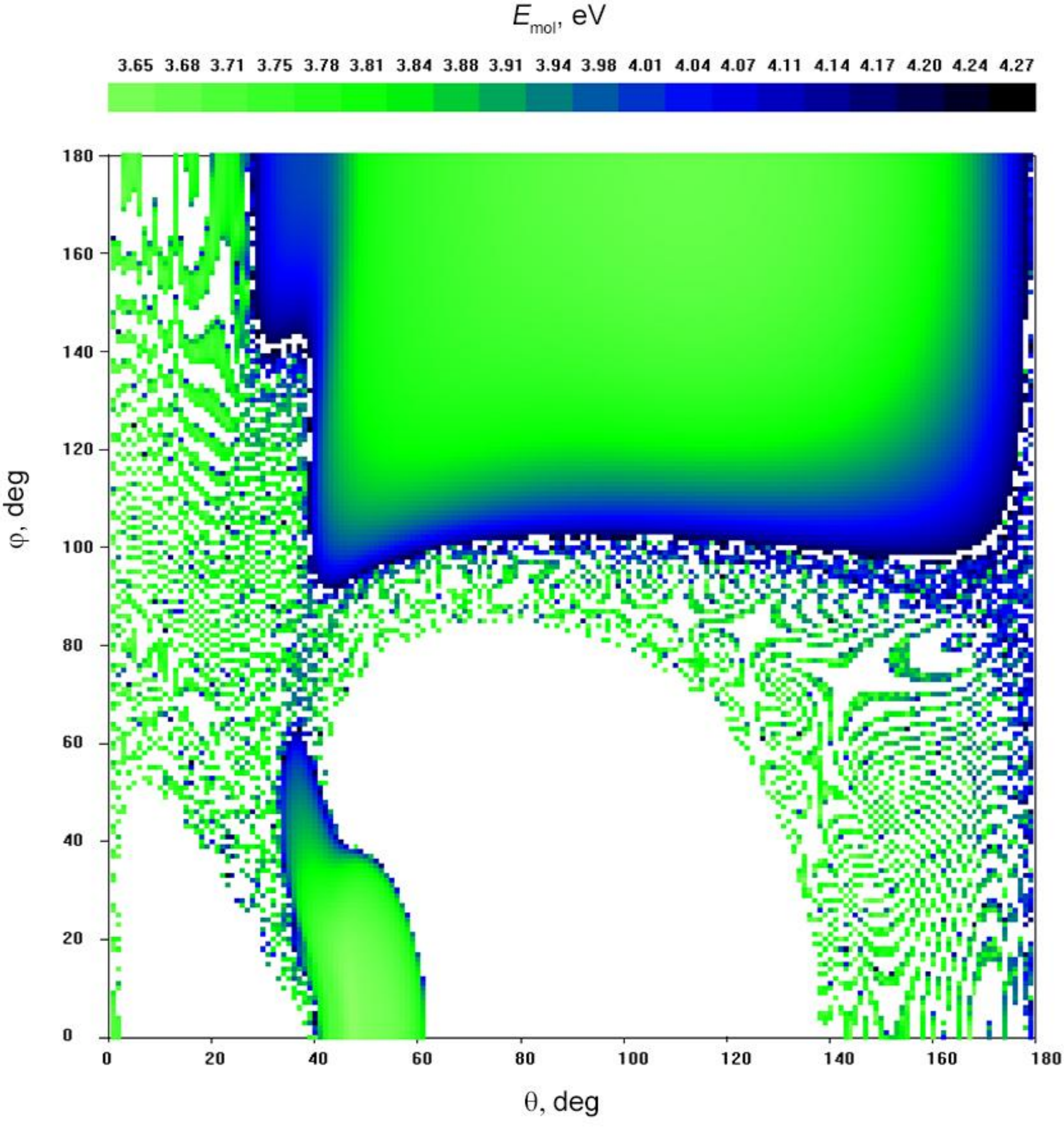

**Fig. 3** for the paper by *V. M. Akimov et al.* “The existence regions for transcomplex recombination of heavy ions. Two-dimensional recombination regions and surroundings indices”

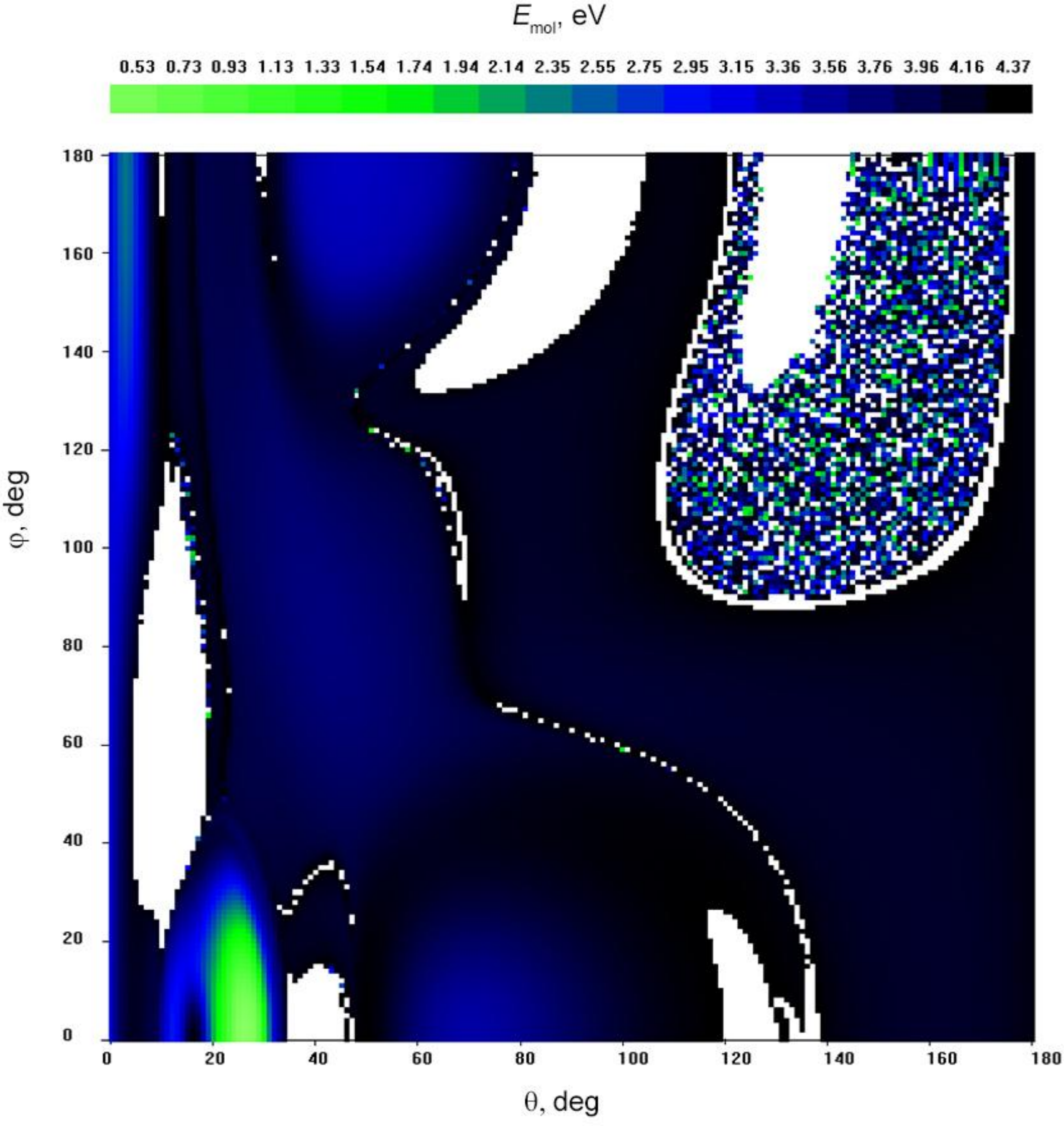

**Fig. 4** for the paper by *V. M. Akimov et al.* "The existence regions for transcomplex recombination of heavy ions. Two-dimensional recombination regions and surroundings indices"

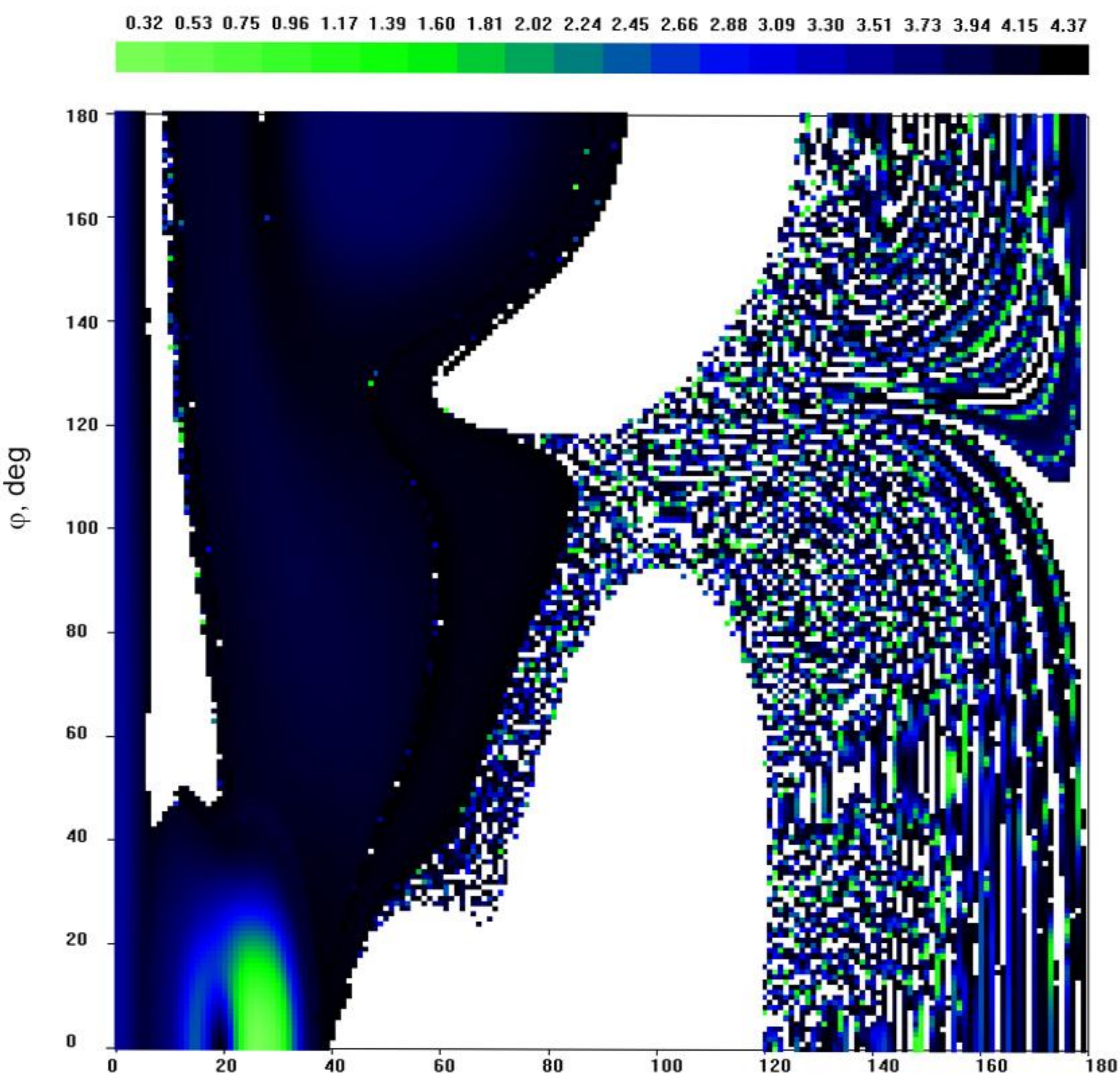

**Fig. 5** for the paper by *V. M. Akimov et al.* “The existence regions for transcomplex recombination of heavy ions. Two-dimensional recombination regions and surroundings indices”

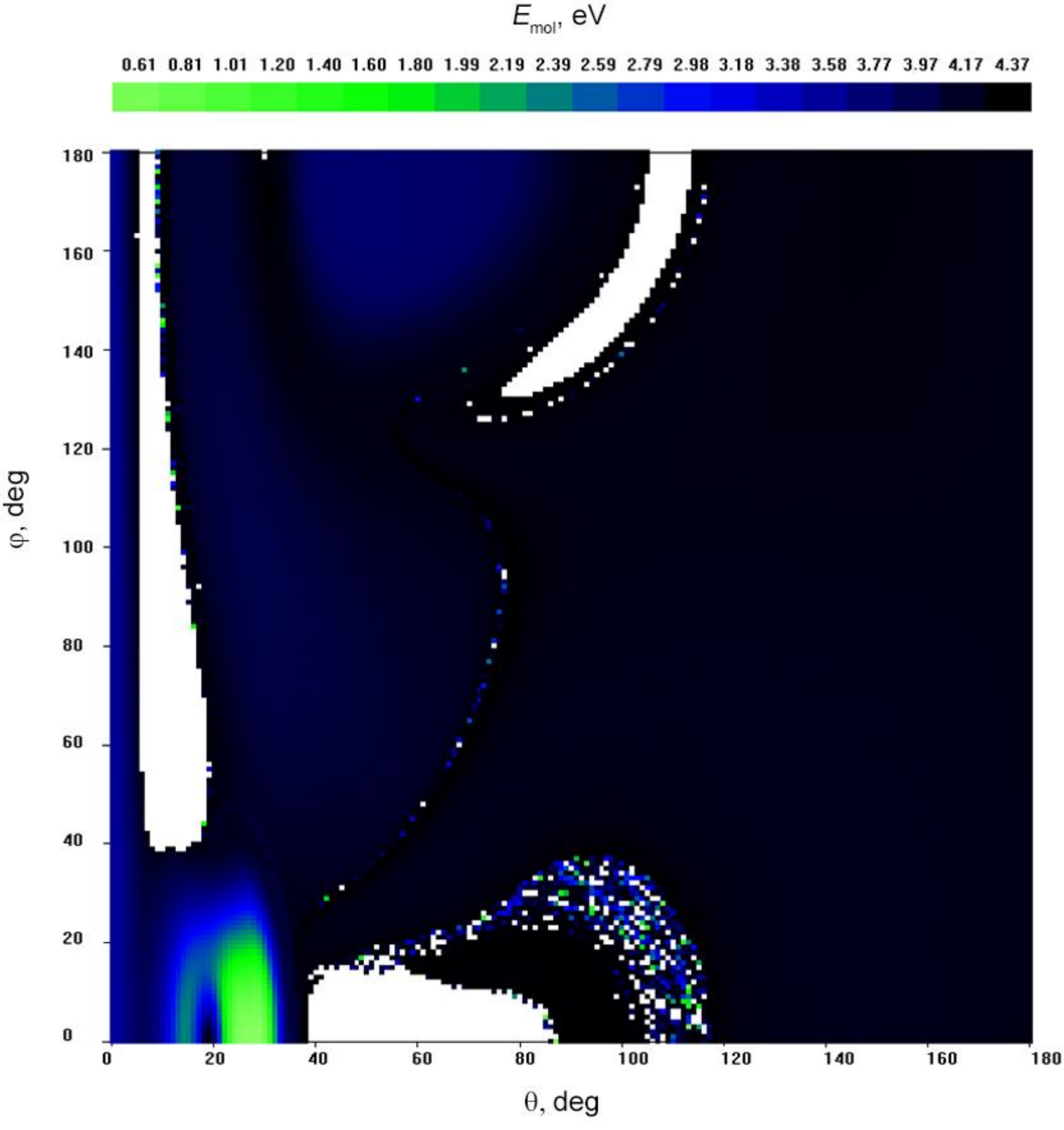

**Fig. 6** for the paper by *V. M. Akimov et al.* "The existence regions for transcomplex recombination of heavy ions. Two-dimensional recombination regions and surroundings indices"

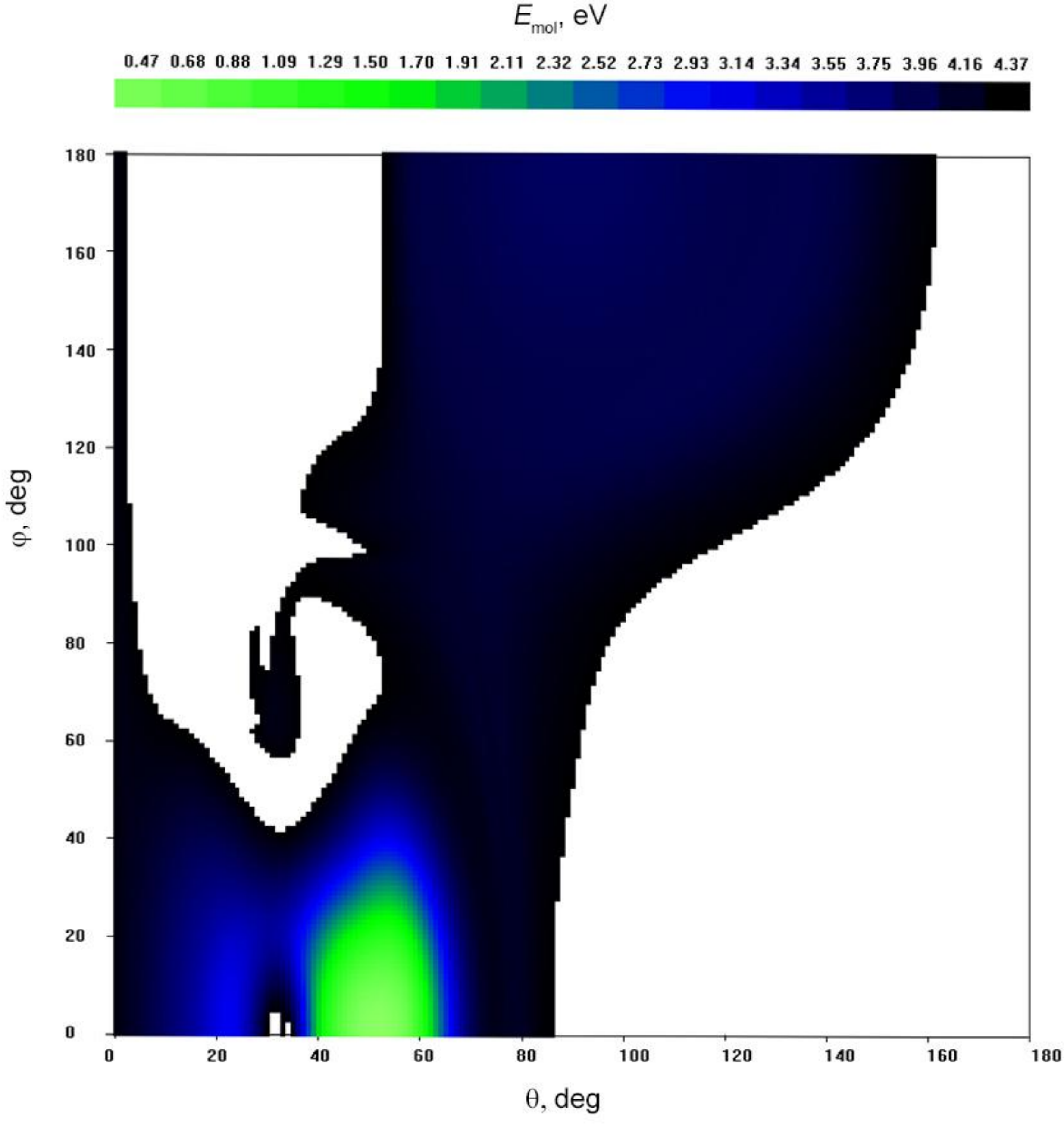

**Fig. 7** for the paper by *V. M. Akimov et al.* "The existence regions for transcomplex recombination of heavy ions. Two-dimensional recombination regions and surroundings indices"

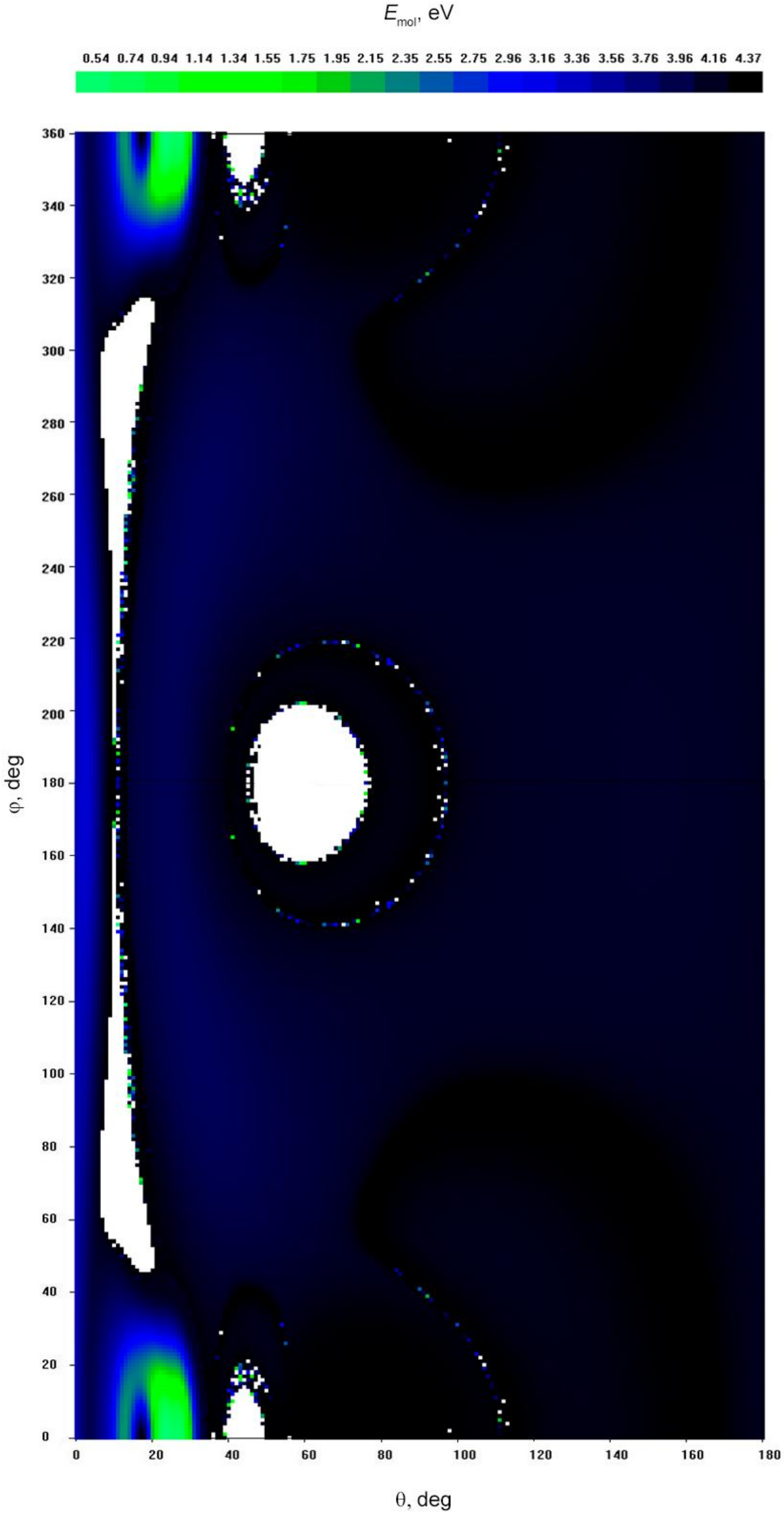

**Fig. 8** for the paper by *V. M. Akimov et al.* "The existence regions for transcomplex recombination of heavy ions. Two-dimensional recombination regions and surroundings indices"

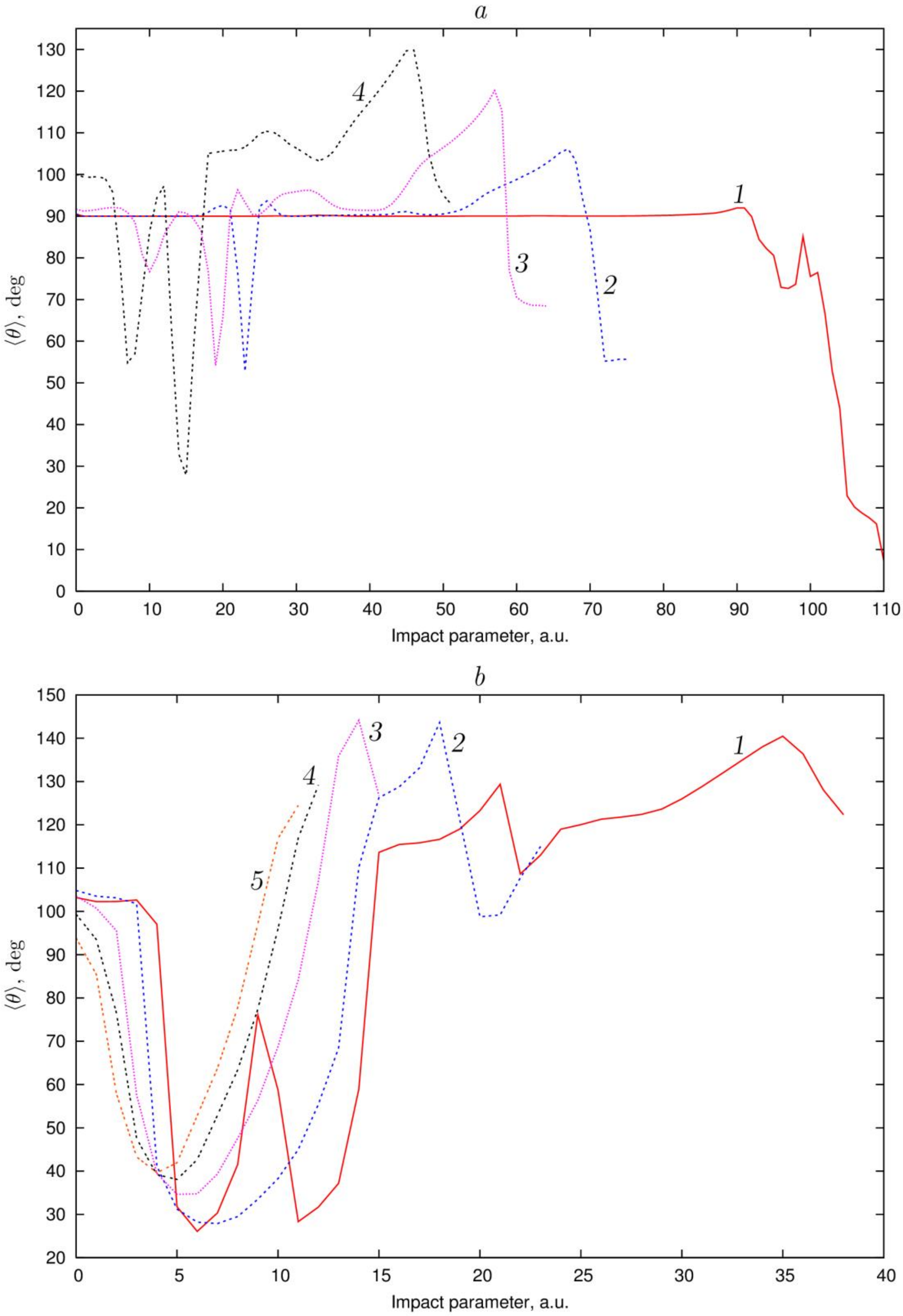

**Fig. 9** for the paper by *V. M. Akimov et al.* "The existence regions for transcomplex recombination of heavy ions. Two-dimensional recombination regions and surroundings indices"

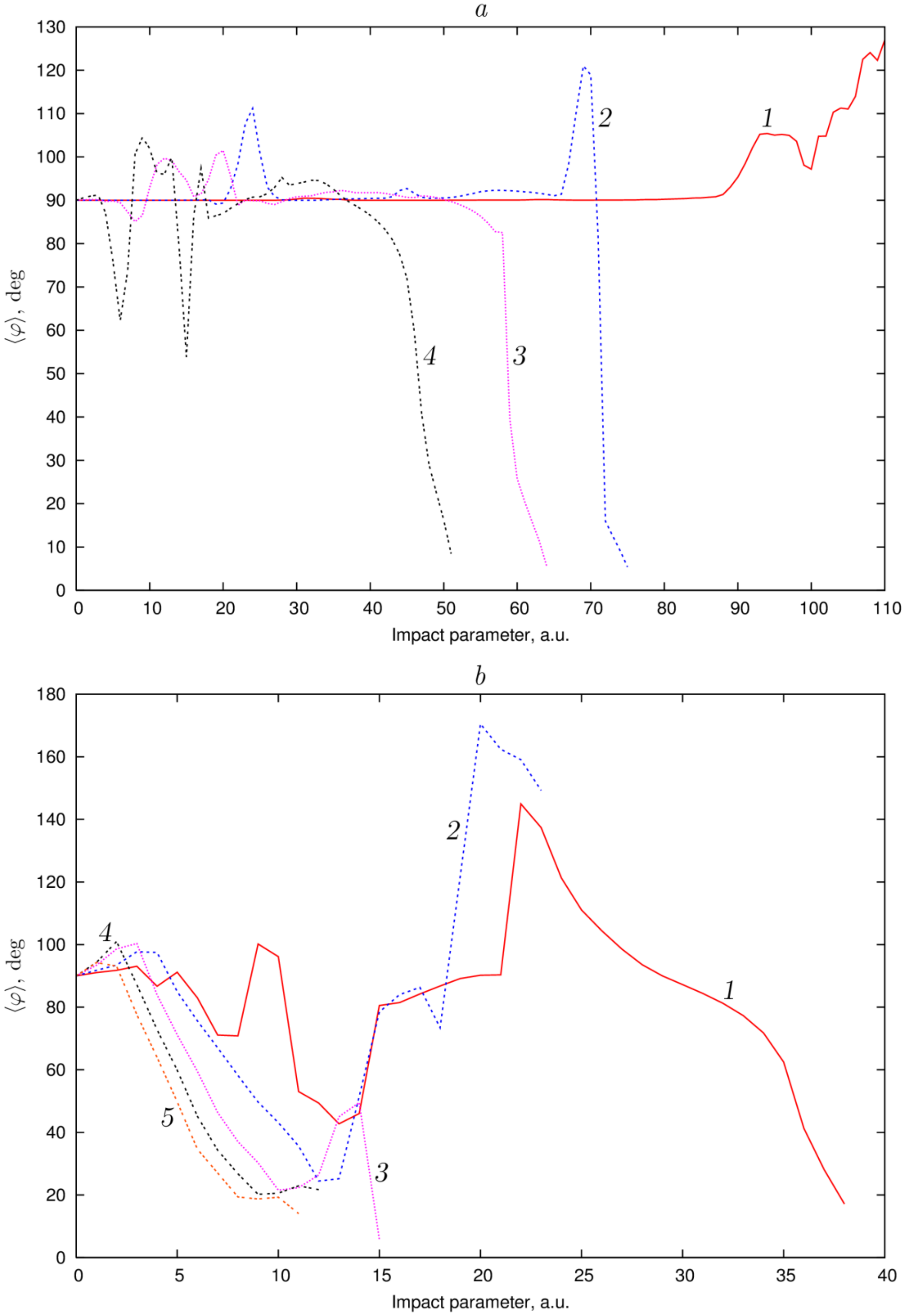

**Fig. 10** for the paper by *V. M. Akimov et al.* "The existence regions for transcomplex recombination of heavy ions. Two-dimensional recombination regions and surroundings indices"

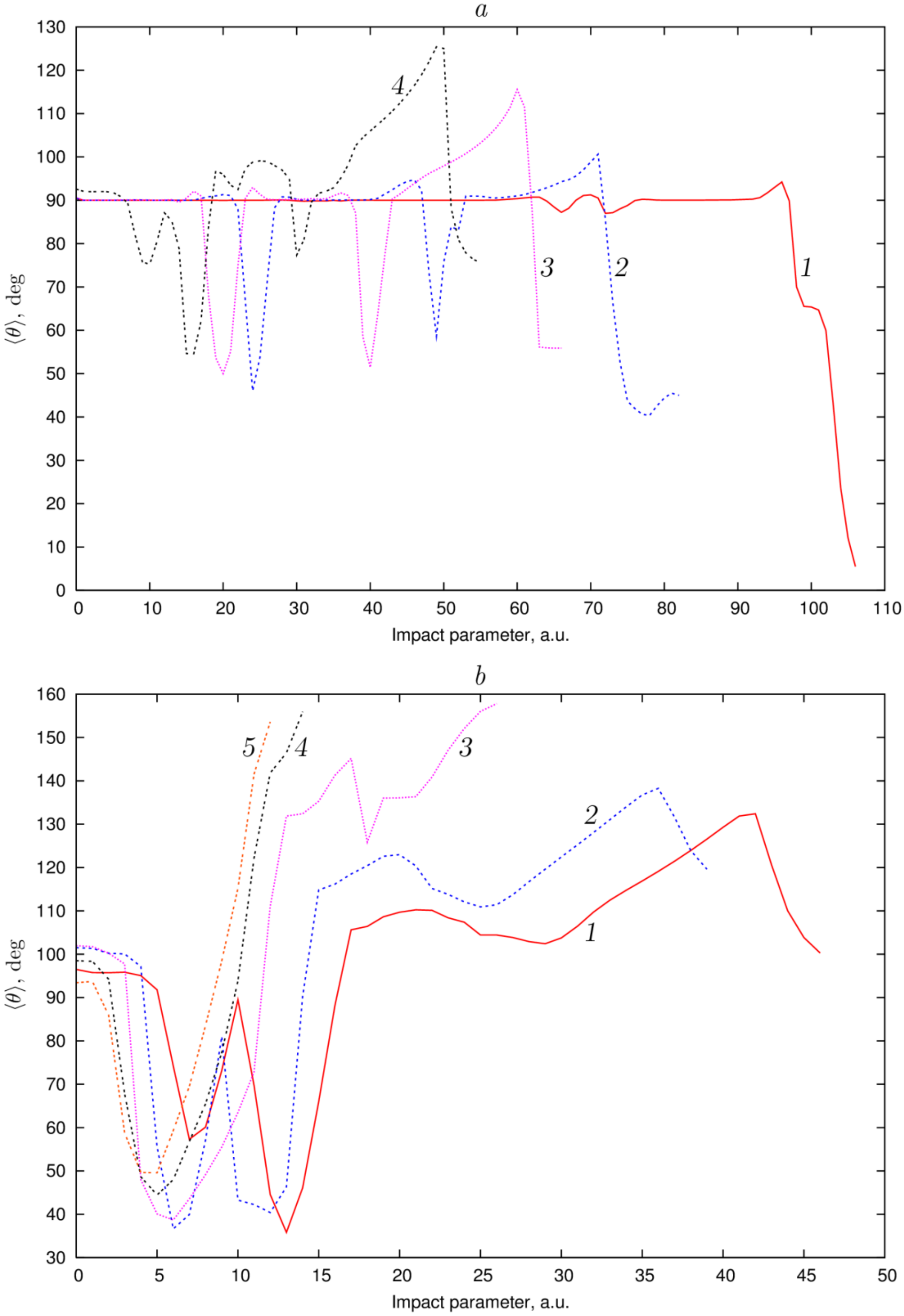

**Fig. 11** for the paper by *V. M. Akimov et al.* "The existence regions for transcomplex recombination of heavy ions. Two-dimensional recombination regions and surroundings indices"

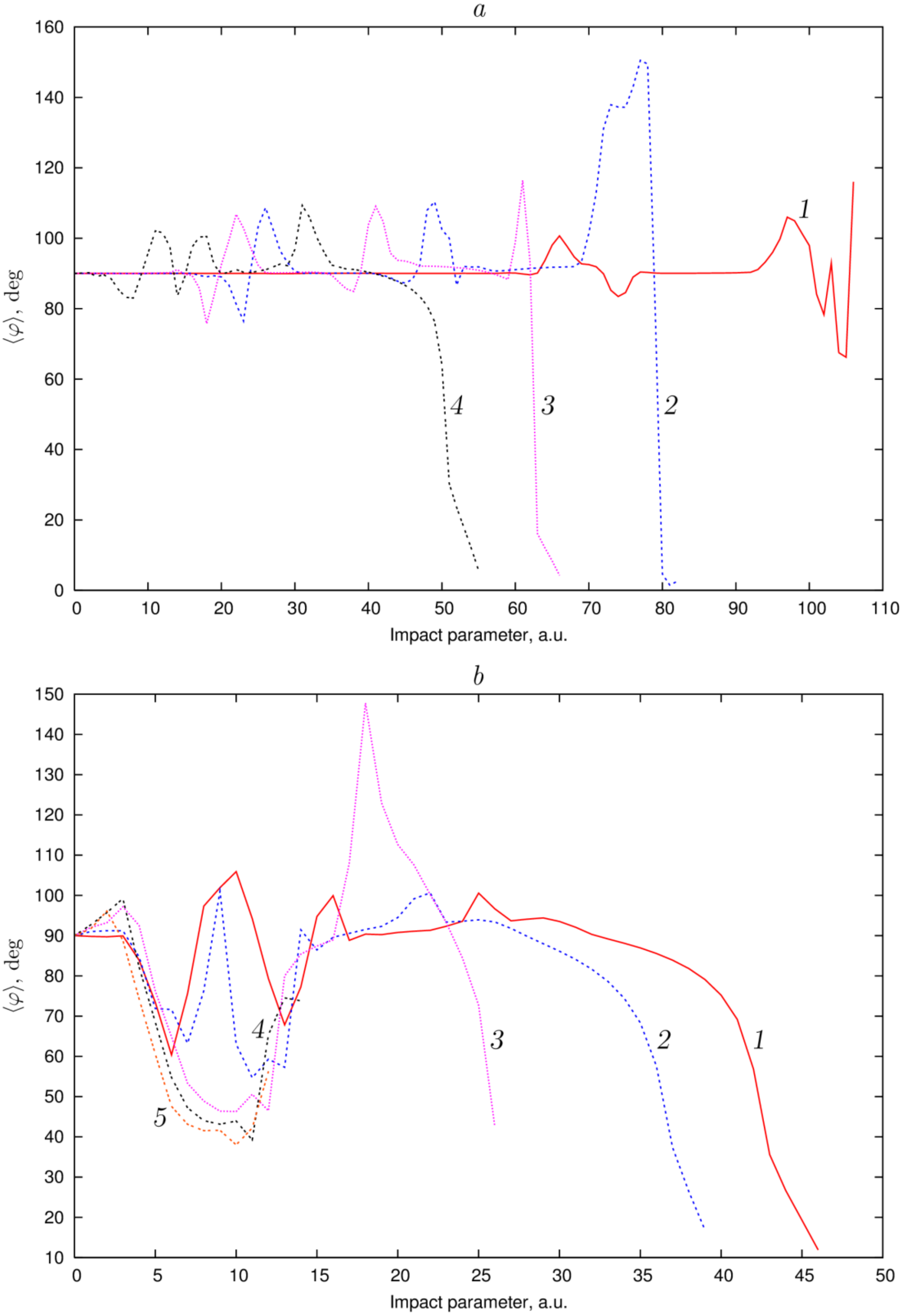

**Fig. 12** for the paper by *V. M. Akimov et al.* "The existence regions for transcomplex recombination of heavy ions. Two-dimensional recombination regions and surroundings indices"

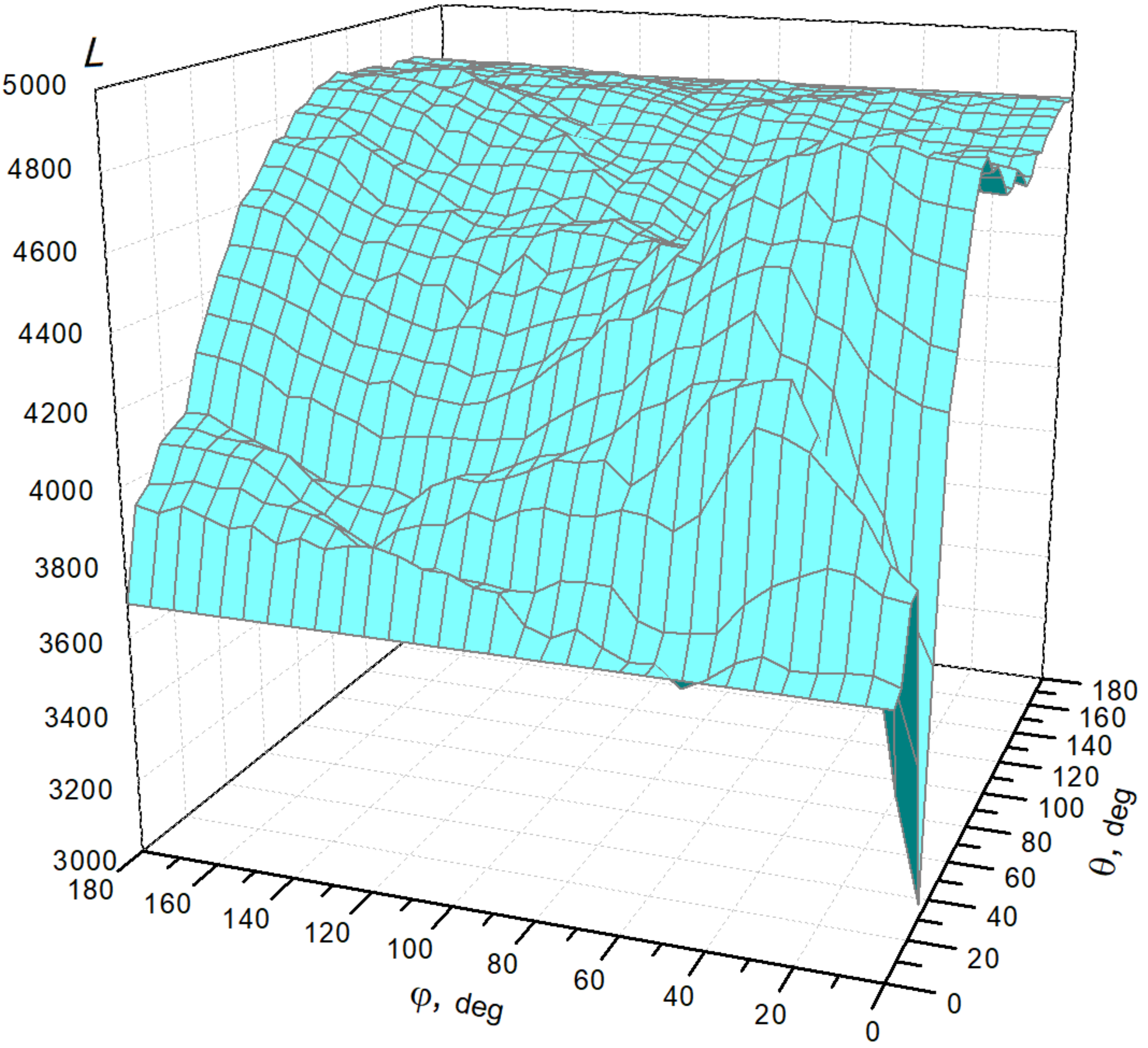

**Fig. 13** for the paper by *V. M. Akimov et al.* "The existence regions for transcomplex recombination of heavy ions. Two-dimensional recombination regions and surroundings indices"

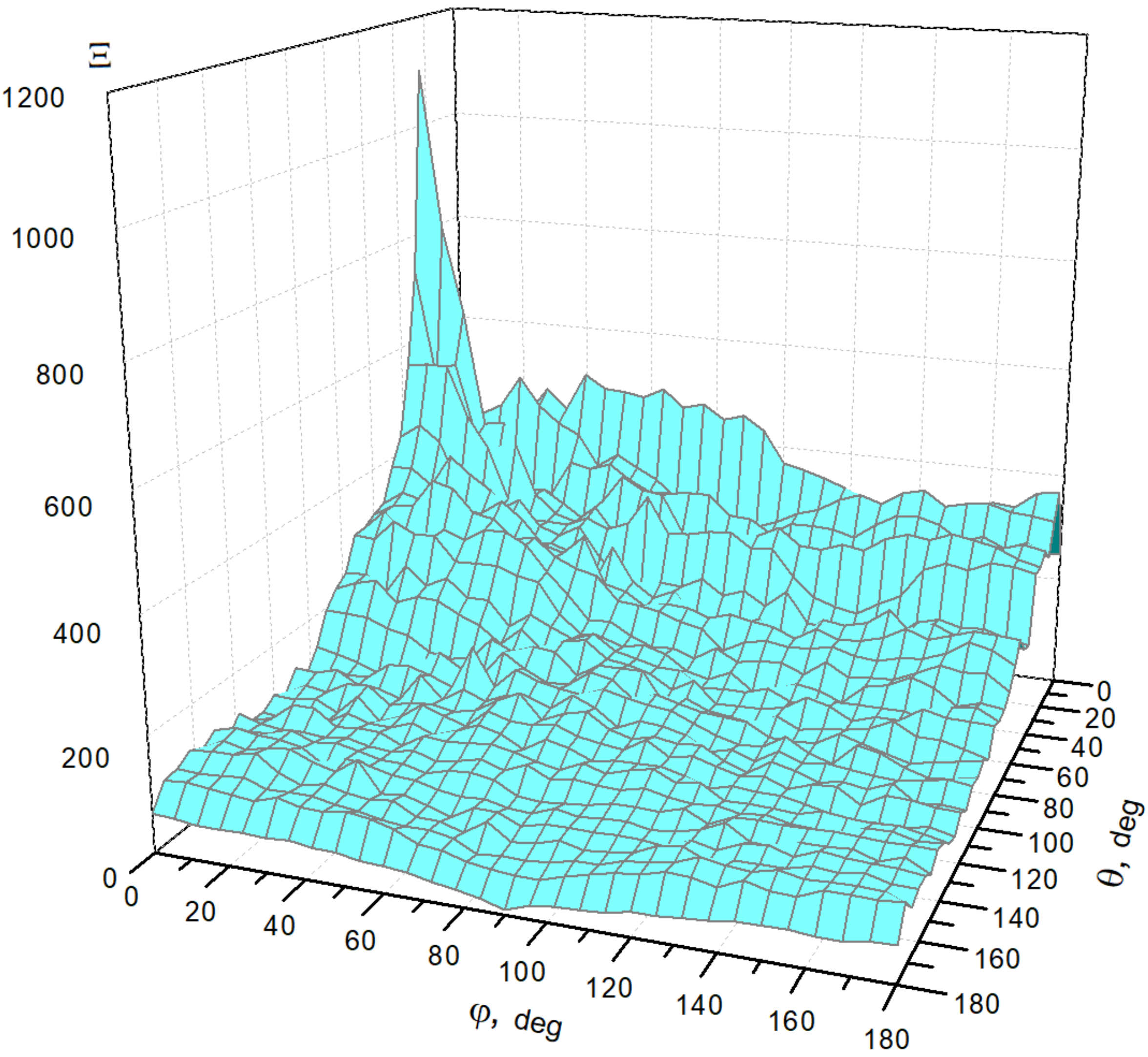

**Fig. 14** for the paper by *V. M. Akimov et al.* “The existence regions for transcomplex recombination of heavy ions. Two-dimensional recombination regions and surroundings indices”